\documentclass[preprint]{aastex}

\slugcomment{\begin{minipage}[t]{4cm}
\begin{flushright}
UMN-TH-2411/05\\
FTPI-MINN-05/38\\
astro-ph/0509183 \\
August 2005 
\end{flushright}
\end{minipage}}

\begin{document}

\title{Hierarchical Growth and Cosmic Star Formation:  Enrichment, Outflows and Supernova Rates.}
\author{Fr\'ed\'eric~Daigne}
\affil{Institut d'Astrophysique de Paris, UMR 7095, CNRS, Universit\'e Pierre et Marie Curie-Paris VI, 98 bis bd Arago, F-75014, Paris, France}
\email{daigne@iap.fr}

\author{Keith A. Olive}
\affil{William I. Fine Theoretical Physics Institute, School of Physics and Astronomy,
University of Minnesota, Minneapolis, MN 55455 USA}

\author{Joe Silk}
\affil{Department of Physics, University of Oxford, Keble Road, Oxford OX1 3RH,
and Institut d'Astrophysique, 98 bis Boulevard Arago, Paris 75014, France}

\author{Felix Stoehr}
\affil{Institut d'Astrophysique de Paris, UMR 7095, CNRS, Universit\'e Pierre et Marie Curie-Paris VI, 98 bis bd Arago, F-75014, Paris, France}

\and

\author{Elisabeth Vangioni}
\affil{Institut d'Astrophysique de Paris, UMR 7095, CNRS, Universit\'e Pierre et Marie Curie-Paris VI, 98 bis bd Arago, F-75014, Paris, France}

\begin{abstract}
The
cosmic star formation histories are evaluated for  different minimum masses of
the initial halo structures, 
with allowance for
realistic gas outflows.
With a minimum halo mass of $10^{7}$--$10^{8}\
\mathrm{M}_{\odot}$  and 
a moderate outflow efficiency, we 
 reproduce both the current baryon fraction 
 and the early chemical enrichment of the
IGM. The intensity of the formation rate of ``normal'' stars is also
well constrained by the observations: it has to be dominated by 
star formation in elliptical galaxies, except perhaps  at very low
redshift. The fraction of baryons in stars is  predicted as are also
the type Ia and II supernova event
rates. Comparison with SN observations in the redshift range 
$z=0-2$ allows us to set  strong constraints on the time delay of type Ia
supernovae (a total delay of $\sim$4 Gyr is required to fit the data), the lower end of the
mass range of the progenitors (2 - 8 $\mathrm{M}_{\odot}$) and  the
fraction of white dwarfs that reproduce the  type Ia supernova (about 1 per
cent). 
The intensity in the initial starburst of zero  metallicity  stars below
270 $\mathrm{M_{\odot}}$
must be 
limited in order to avoid premature overenrichment of the IGM. 
Only about 10 - 20 \% of the metals present in the IGM  at $z = 0$ have been produced
by population III stars at very high $z$. The remaining 80 - 90 \% are ejected later  by
galaxies forming normal stars, with a maximum outflow efficiency 
occurring at a redshift of about 5.
We conclude that
$10^{-3}$ of the  mass in baryons must lie in first  massive stars  
in order to produce  enough ionizing photons
to  allow early reionization of the IGM by  $z\sim 15$. 
\end{abstract}

\section{Introduction}

Chemical evolution is a key to understanding the hybrid role of
massive star formation in the early universe with regard to the epoch
of reionization, the heavy element abundances of the oldest stars and
the high $z$ intergalactic medium, and the mass outflows associated with
galaxy formation.  In addition, predicted supernova rates provide us with an
independent probe of the early epoch of star formation. 
Combining abundance and supernova rate predictions
allows us to develop an improved understanding of both the cosmic star
formation history and of the enrichment of the IGM, as well as to
elucidate the nature of Population III (see e.g. the recent review of \citet{ciardi:05}).

The prevalent view  is  that the first stars encompassed the
mass range 100 to 1000 M$_\odot.$ Recent support for this possibility stems from
the need to reionize the Universe at  high redshift \citep[]{cen:03a,haiman:03,wyithe:03,bromm:04a}
as indicated by the WMAP first-year data \citep{kogut:03}. The possibility for early 
reionization by the first galaxies was considered by \citet{ciardi:03} and
further support for this hypothesis is  based on chemical abundance patterns at low 
metallicity \citep{wasserburg:00,oh:01,qian:01,qian:05}.  

However, the robustness
of the conclusion that very massive stars (VMS) were necessarily present among the 
first stars has been questioned \citep{venkatesan:03b,tumlinson:04,tumlinsona:05}. A ``normal" initial
mass function (IMF) may be capable of producing ionization consistent with WMAP
\citep{venkatesan:03a,wyithe:03}, and within these constraints, it was argued \citep{venkatesan:04} that
a broad set of 
chemical abundances may be better fit using the yields of \citet{umeda:03} for stars with masses
in the 1- 50 M$_\odot$ range than with the yields from pair-instabilty supernovae (PISN) \citep{heger:02}.
Indeed, these results were confirmed in \citet[][hereafter DOSVA]{daigne:04}  where it was argued that
a top-heavy IMF without VMS supplied a better fit to low metallicity abundance data
while still accounting for the early re-ionization of the Universe
when using a detailed model of cosmic chemical evolution. Furthermore, when VMS
were assumed to be the predominant source for Pop III, it was found that there could not
be enough of these stars
to have accounted for reionization without producing anomalous chemical
abundance ratios in both old halo stars and in the IGM.
Even partial suppression of the yields of pair instability SNe via
rotation and fall-back should not vitiate this conclusion in view of
the anomalous abundance ratios and the large discrepancy  found by DOSVA.

In this paper, we improve on an earlier model for primordial star
formation and chemical evolution (DOVSA)
by using a more realistic 
model of structure formation to study the role of baryonic
outflows and the enrichment of the IGM. Our principal result is that only if
Population III predominantly consisted of stars in the ``normal'' mass
range, but with a top-heavy IMF, that is to say dominated by stars in
the mass range 40 to 100 M$_\odot,$ can one simultaneously account for
reionization and chemical enrichment, both of galaxies and of the
IGM. However, this mass range is not capable of sufficient Si production
and therefore, more massive stars may be necessary to account for the 
inferred abundance of Si in the IGM.
Such a bimodal approach to star formation allows us to account
both for the cosmic star formation history and global chemical evolution of
the universe. The predicted supernova rates agree with those observed.
We also infer a baryonic outflow rate that allows us to reconcile the
observed baryon fraction in massive galaxies with that observed in
clusters, the CMB and the Lyman alpha forest, as well as that
predicted by primordial nucleosynthesis.

The outline of this paper is as follows. In \S~\ref{sec:method} we
present the details of the chemical evolution model, especially the
treatment of the structure formation and of the galactic winds. We also outline
the key parameters used in the models which affect our conclusions. In
\S~\ref{sec:model0}, we discuss the normal mode of star formation and
we show that the minimum masses of the halos of star forming structures
as well as the efficiency of the galactic winds are well constrained
by the observations. This allows us to select a few realistic models. In
\S~\ref{sec:SNae}, we compute the expected SNII and SNIa rates in these
models and deduce the time delay inferred for SNIa to fit the data.
\S~\ref{sec:popIII} is devoted to the addition of an initial starburst
of massive stars in our scenario. We summarize our results in
\S~\ref{sec:conclusions}.

 
\section{Hierarchical formation scenario and associated galactic winds}
\label{sec:method}

\subsection{A cosmic evolutionary model}

The chemical evolution model used in this paper has been described in DOVSA. 
It is a generalized version of standard models designed to follow one specific structure such as the Milky Way (for a review, see \citet{tinsley:80}). We describe baryons in the Universe by two large reservoirs. The first  is associated with collapsed structures (hereafter the ``structures'') and is divided in two sub-reservoirs: the gas (hereafter the ``interstellar medium'' or ISM) and the stars and their remnants (hereafter the ``stars''). The second reservoir corresponds to the medium in between the collapsed structures (hereafter the ``intergalactic medium'' or IGM). The evolution of the baryonic mass of these reservoirs, i.e.  $M_\mathrm{IGM}(t)$ of the IGM, $M_\mathrm{ISM}(t)$ of the ISM and $M_{*}(t)$ of the stars  is governed by a set of differential equations (see section 2 in DOVSA):
\begin{equation}
\frac{dM_\mathrm{IGM}}{dt}=-\frac{dM_\mathrm{struct}}{dt}=-a_\mathrm{b}(t)+o(t),
\end{equation}

\begin{equation}
\frac{dM_\mathrm{*}}{dt}=\Psi(t)-e(t)\ \mathrm{and}\ \frac{dM_\mathrm{ISM}}{dt}=\frac{dM_\mathrm{struct}}{dt}-\frac{dM_\mathrm{*}}{dt}\ .
\end{equation}

\begin{figure}
\begin{center}
\resizebox{\textwidth}{!}{\includegraphics{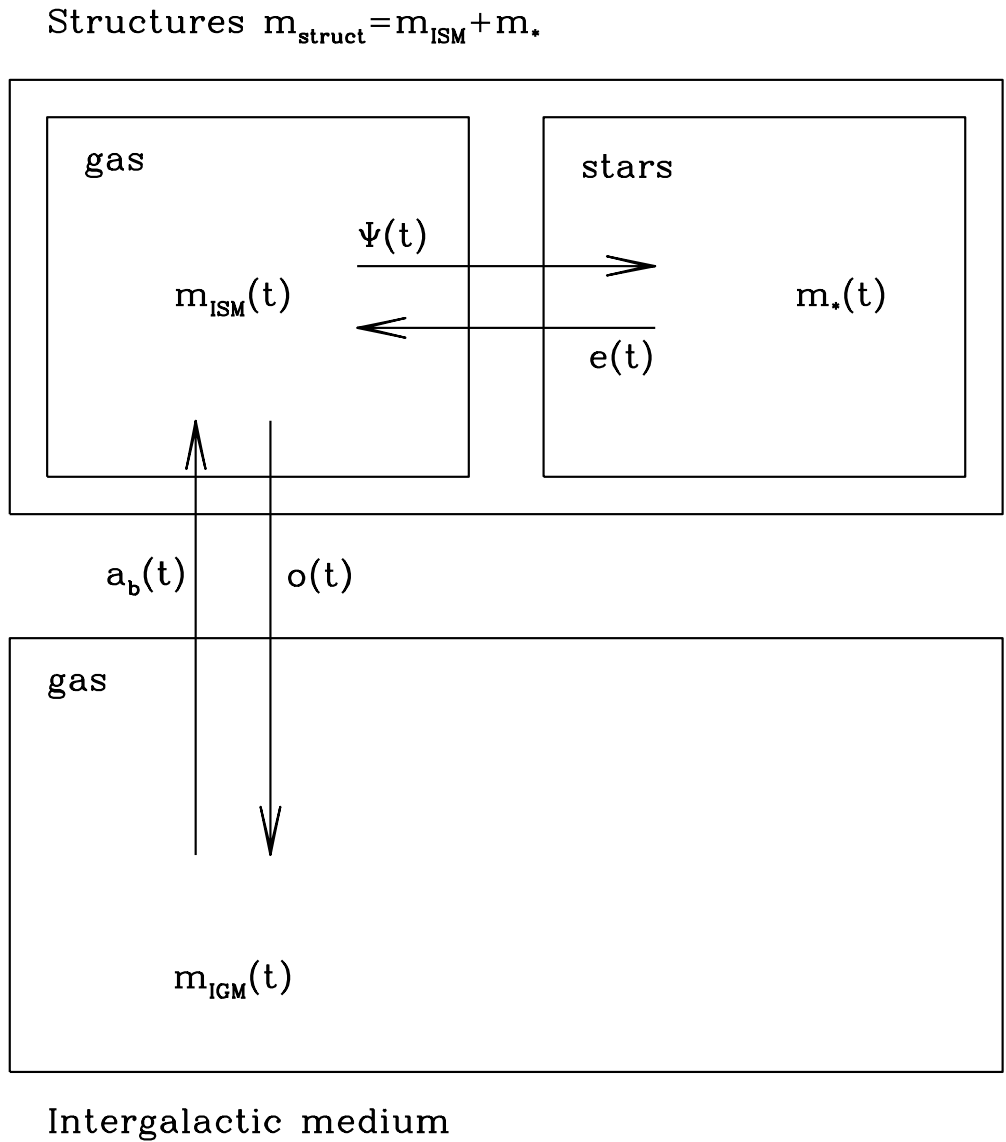}}
\end{center}
\caption{A sketch of our cosmic evolutionary model. The three baryon reservoirs (IGM, ISM and stars) exchange mass via four physical processes (structure formation, $a_\mathrm{b}$, star formation, $\Psi$, gas ejection by stars, $e$, outflow from the structures, $o$).}
\label{fig:schemacode}
\end{figure}

\noindent In addition, we have $M_\mathrm{ISM}(t)+M_\mathrm{*}(t)=M_\mathrm{struct}(t)$, corresponding to the total baryonic mass of the structures, and $M_\mathrm{IGM}(t)+M_\mathrm{struct}(t)=\mathrm{constant}$, which is the total baryonic mass of the Universe. As can be seen, these equations are controlled by four rates which represent four fundamental processes (see sketch in Figure~\ref{fig:schemacode}): the formation of structures through the accretion of baryons from the IGM,
$a_\mathrm{b}(t)$; the formation of stars  through the transfer of baryons from the ISM, $\Psi(t)$; the ejection of enriched gas by stars, $e(t)$ and the outflow of baryons from the structures into the IGM, 
$o(t)$.

We track the chemical composition of the ISM and the IGM separately as a function of time (or redshift). The differential equations governing the evolution of the mass fraction $X_{i}^\mathrm{ISM}$ ($X_{i}^\mathrm{IGM}$) of element $i$ in the ISM (IGM) are given by equations (6) and (7) in section 2 in DOVSA. In addition, we also compute the ionizing UV fluxes from the stars (equation (12) in section 3 in DOVSA) and the rate of explosive events (type Ia  and gravitational collapse supernovae).\\

The age $t$ of the Universe which appears in the equations is related to the redshift by
\begin{equation}
\frac{dt}{dz}=\frac{9.78 h^{-1}\ \mathrm{Gyr}}{(1+z)\sqrt{\Omega_\mathrm{\Lambda}+\Omega_\mathrm{m}\left(1+z\right)^{3}}}\ ,
\end{equation}
assuming the cosmological parameters of the so-called ``concordance model'', with a density of matter $\Omega_\mathrm{m}=0.27$ and a density of ``dark energy'' $\Omega_\mathrm{\Lambda}=0.73$ (Spergel et al. 2003), and taking $H_{0}=71\ \mathrm{km/s/Mpc}$ for the Hubble constant ($h=0.71$). This allows us to trace all the quantities we describe as a function of redshift. The input stellar data (lifetimes, mass and type of the remnant, metal yields, UV flux) are taken to be dependent both of the mass and the metallicity of the star (see DOVSA for more details).\\

Building upon our original model, we have included the following improvements: (i) the baryon accretion rate, $a_\mathrm{b}(t)$,  by the structures is now computed in the framework of the hierarchical scenario of structure formation, \citep{press:74,sheth:99,jenkins:01,wyithe:03b}. This is described in section~\ref{sec:ab}. (ii) the baryon outflow rate, $o(t)$, from the structures now includes a redshift-dependent efficiency, which accounts for the increasing escape velocity of the structures as the galaxy assembly is in progress. This is described in section~\ref{sec:o}. Note that in DOVSA, the outflow contained two sources: $o(t)=o_\mathrm{w}(t)+o_\mathrm{SN}(t)$ where $o_\mathrm{w}(t)$ is a global outflow powered by stellar explosions (galactic winds) and $o_\mathrm{SN}(t)$ corresponds to stellar supernova ejecta that are flushed directly out of the structures. Since it was shown in DOVSA that the second term has a very small effect on the results, we have neglected it in this work (in DOVSA's notation, this is equivalent to setting  $\alpha=0$).

\subsection{Hierarchical formation scenario}
\label{sec:ab}
As seen in DOVSA, the term $a_\mathrm{b}(t)$ which stands for the process of structure formation can have a strong impact on the high redshift evolution, intensity of the star formation process, ionizing flux, and the enrichment of the IGM, as it governs the size of the reservoir of baryons available for star formation. In DOVSA, although this term was estimated in an oversimplified way, we assumed only that structure formation efficiency decays exponentially. This first attempt allowed us to point out the key role of this term in the generalized chemical evolution model we have implemented. Here, we improve the model with 
a more realistic description of structure formation. 

We adopt the framework of the hierarchical scenario where small structures are formed first. At redshift $z$, the comoving density of dark matter halos in the mass range $[M,M+dM]$ is $f_\mathrm{PS}(M,z)dM$, with 
\begin{equation}
\int_{0}^{\infty} dM\ M f_\mathrm{PS}(M,z) = \rho_\mathrm{DM}\ ,
\end{equation}
where $\rho_\mathrm{DM}$ is the comoving dark matter density. The distribution function of halos $f_\mathrm{PS}(M,z)$ is computed using the method described in \citet{jenkins:01} using a code provided by A. Jenkins. It follows  the standard theory \citep{press:74}, including the modification of \citet{sheth:99} and assumes a primordial power spectrum with a power-law index $n=1$ and the fitting formula to the exact transfer function for non-baryonic cold dark matter given by \citet{bond:84}. We adopt a rms amplitude $\sigma_{8}=0.9$ for mass density fluctuations in a sphere of radius $8\ h^{-1}\ \mathrm{Mpc}$.\\

We assume that the baryon distribution traces the dark matter distribution without any bias so that the density of baryons is just proportional to the density of dark matter by a factor $\Omega_{b}/\left(\Omega_{m}-\Omega_{b}\right)$. We take a baryonic density $\Omega_\mathrm{b}=0.044$ \citep{spergel:03}. We parametrize the fact that stars can form only in structures which are suitably dense by defining the minimum mass $M_\mathrm{min}$ of a dark matter halo of the collapsed structures where star formation occurs. This mass could be related in principle with the critical temperature at which the cooling processes become efficient enough to allow star formation. In fact, this critical temperature, and hence the minimum mass $M_\mathrm{min}$, should evolve with redshift $z$ as the cooling processes of the hot gas in structures depend strongly on the chemical composition and ionizing state of the gas. It is however beyond the scope of this study to include such a detailed analysis so we prefer to keep $M_\mathrm{min}$ constant and to consider it as a free parameter of the model. The fraction of baryons at redshift $z$ which are in such structures is then given by
\begin{equation}
f_\mathrm{b,struct}(z) = \frac{\int_{M_\mathrm{min}}^{\infty} dM\ M f_\mathrm{PS}(M,z)}{\int_{0}^{\infty} dM\ M f_\mathrm{PS}(M,z)}\ .
\label{eq:fb}
\end{equation}
Therefore, the mass flux $a_\mathrm{b}$ can be estimated by
\begin{eqnarray}
a_\mathrm{b}(t) & = & \Omega_\mathrm{b}\left(\frac{3H_{0}^{2}}{8\pi G}\right)\ \left(\frac{dt}{dz}\right)^{-1}\ \left|\frac{d f_\mathrm{b,struct}}{dz}\right|\nonumber\\
&  = & 1.2 h^{3}\ \mathrm{M_{\odot}/yr/Mpc^{3}}\ \left(\frac{\Omega_\mathrm{b}}{0.044}\right)\left(1+z\right)\sqrt{\Omega_\mathrm{\Lambda}+\Omega_\mathrm{m}\left(1+z\right)^{3}}\left|\frac{df_\mathrm{b,struct}}{dz}\right|\ .
\end{eqnarray}
This is the new expression of $a_\mathrm{b}(t)$ used in the model. Even if it represents a significative improvement compared to the treatment of the same term in DOVSA, one should keep in mind that this derivation is only partially consistent, as the actual fraction of baryons in collapsed structures with dark matter halos of mass $M\ge M_\mathrm{min}$ will differ from the value given by equation~\ref{eq:fb} due to the outflow of baryons from the star-forming structures towards the IGM. However, since $o(t)$ is always small compared to $a_\mathrm{b}(t)$, this is a small correction for most of the evolutionary history.\\

Finally, this formalism allows us to indirectly fix the initial redshift $z_\mathrm{init}$ where the first stars form in the simulation. This redshift is a priori smaller than the redshift $z_\mathrm{PS}(M_\mathrm{min})$ at which the first dark matter halos of mass $M_\mathrm{min}$ appear, as the star formation process can be delayed due to the time necessary to cool the collapsing gas. Instead of fixing the value of $z_\mathrm{init}$, we prefer to fix the size of the gas reservoir when the star formation starts, i.e. the initial fraction $f_\mathrm{b,struct}(z_\mathrm{init})$ of baryons in collapsed structures which form stars. We usually adopt $f_\mathrm{b,struct}(z_\mathrm{init})=1\ \%$ but we also consider $0.1\ \%$ and $5\ \%$ for some specific cases. Supernova feedback generically gives a star formation efficiency of 1 or 2 percent per dynamical time \citep{silk:03} and an inefficiency of this order is inferred in  essentially all current epoch  star-forming galaxies \citep{kenni:03}.

 The corresponding initial redshift $z_\mathrm{init}$ is computed from equation~\ref{eq:fb} for a given minimum mass $M_\mathrm{min}$ (see table~\ref{tab:model0}). Note that this redshift is usually smaller than that found in hydrodynamical simulations of collapsing gas for the redshift of formation of the first zero-metallicity stars (population III stars): $z\sim 20-30$ \citep{bromm:04b,yoshida:04,bromm:04a}. However, the first stars formed in these simulations are found to form in collapsed  structures with dark matter halos of mass $M\sim 10^{6}\ \mathrm{M_{\odot}}$ which represents only a very small fraction ($\la 10^{-3}$) of the baryons in the Universe at these redshifts. Therefore, this very primordial star formation activity is completely negligible in comparison with the global evolution of the Universe.

\subsection{Outflows}
\label{sec:o}
\begin{figure}
\begin{center}
\resizebox{\textwidth}{!}{\includegraphics{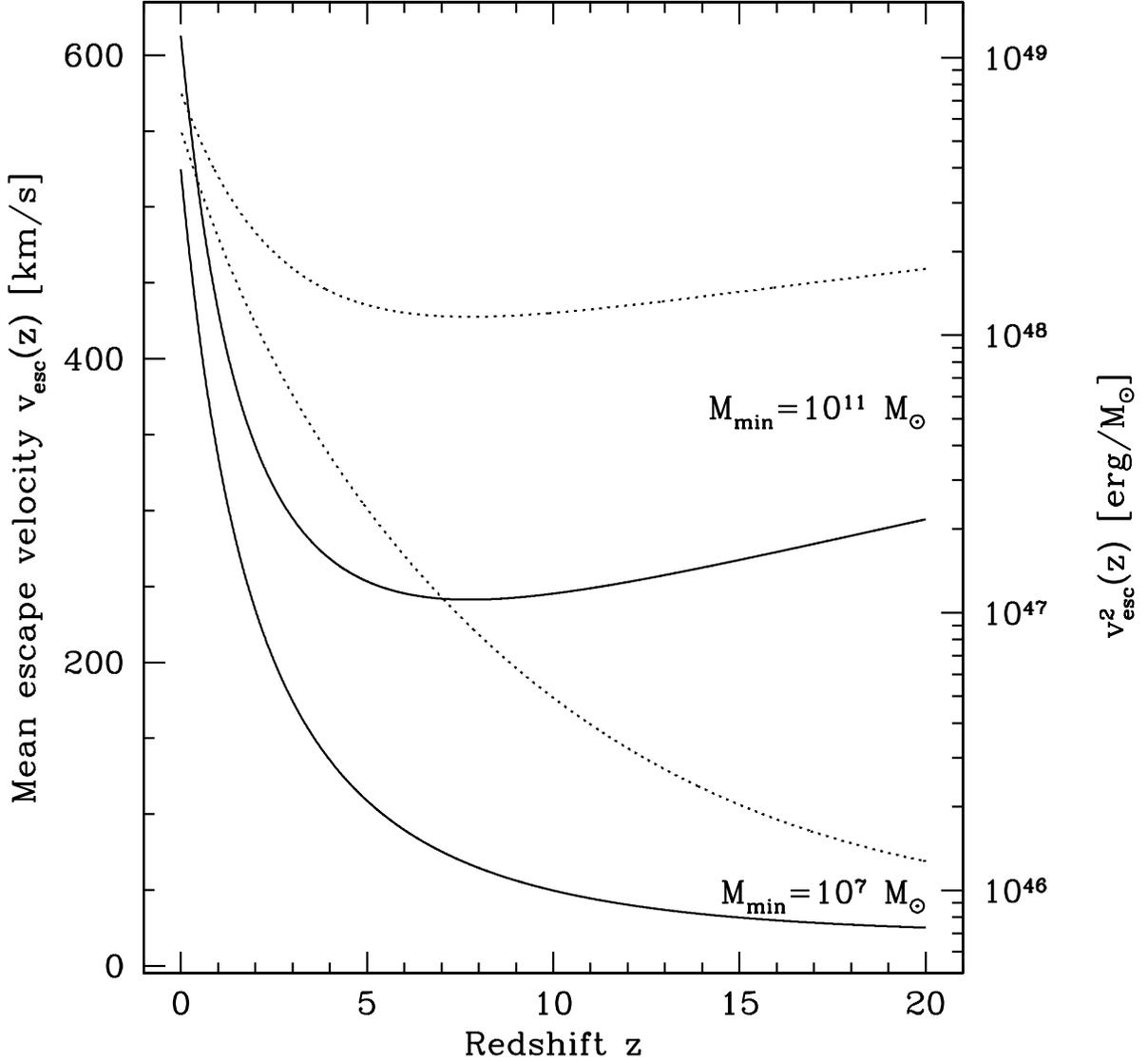}}
\end{center}
\caption{ The averaged escape velocity from structures, $v_\mathrm{esc}$, as a function of redshift is shown for two choices of $M_\mathrm{min}$.
The solid curves correspond to the velocity in units of km/s as indicated by the scale on the left, 
whereas, the dotted curves correspond to $v^2$ in erg/M$_\odot$ with values indicated on the right.}
\label{fig:velocity}
\end{figure}

We compute the outflow by
\begin{equation}
o(t) = \frac{2\epsilon}{v_\mathrm{esc}^{2}(z)}\int_{\max{\left(8\ \mathrm{M_{\odot}},m_\mathrm{d}(t)\right)}}^{m_\mathrm{up}} dm\ \Phi(m)\Psi\left(t-\tau(m)\right) E_\mathrm{kin}(m)\ ,
\end{equation}
where $\Phi(m)$ is the initial mass function (IMF) of stars, $\tau(m)$ is the lifetime of a star of mass $m$ and $m_\mathrm{d}$ is the mass of stars that die at age $t$, $E_\mathrm{kin}(m)$ is the kinetic energy released by the explosion of a star of mass $m$, $\epsilon$ is the fraction of the kinetic energy of supernovae that is available to power the outflow and $v_\mathrm{esc}^{2}(z)$ is the mean square of the escape velocity of structures at redshift $z$ \citep[see e.g.][]{scully:97}. The escape velocity is obtained by mass-averaging the gravitational potential over the mass distribution at redshift $z$:
\begin{equation}
v_\mathrm{esc}^{2}(z) = \frac{\int_{M_\mathrm{min}}^{\infty} dM\ f_\mathrm{PS}(M,z) M \left(2 G M / R\right)}{\int_{M_\mathrm{min}}^{\infty} dM\ f_\mathrm{PS}(M,z) M}\ ,
\end{equation}
where $R$ is the radius of a dark matter halo of mass $M$. The resulting velocity is plotted in Figure~\ref{fig:velocity} for different minimum masses $M_\mathrm{min}$. Notice the strong evolution with $z$: outflows at high redshift are much more efficient and a large fraction (if not all) of the gas in the structures is ejected and mixed in the intergalactic medium, thus contributing to the early enrichment of the IGM.

The total efficiency of this outflow process will depend both on $\epsilon$ and $v_\mathrm{esc}^{2}(z)$ and will therefore vary with redshift. An approximation of the efficiency can be obtained in the ``instantaneous  recycling limit'' where the lifetime of massive stars is neglected by comparing the instantaneous outflow rate with the star formation rate. In this case we have 
\begin{equation}
r_\mathrm{outflow} = \frac{o(t)}{\Psi(t)} \simeq \frac{2\epsilon\ \bar{e}_\mathrm{kin}}{v_\mathrm{esc}^{2}(z)} ,
\end{equation}
where $\bar{e}_\mathrm{kin}$ is the kinetic energy per stellar mass, averaged over the initial mass function:
\begin{equation}
\bar{e}_\mathrm{kin} = \int_{m_\mathrm{inf}}^{m_\mathrm{up}} dm\ \Phi(m) E_\mathrm{kin}(m)\ .
\end{equation}
For a power-law IMF of slope $-(1+x)$, a minimal mass $m_\mathrm{inf}$ and a constant supernova kinetic energy $E_\mathrm{kin}(m)\simeq \mathrm{constant}$, this kinetic energy per stellar mass can be approximated by
\begin{equation}
\bar{e}_\mathrm{kin} \simeq \frac{x-1}{x}\frac{E_\mathrm{kin}}{m_\mathrm{inf}}\times\left\lbrace\begin{array}{ll}
\left(8\ \mathrm{M_{\odot}} / m_\mathrm{inf}\right)^{-x}\ & \mathrm{if}\ m_\mathrm{inf} < 8\ \mathrm{M}_{\odot}\ \mathrm{(normal\ mode)}\\
1                                                         & \mathrm{if}\ m_\mathrm{inf} \ge 8\ \mathrm{M}_{\odot}\ \mathrm{(massive\ mode)}
\end{array}\right.\ .
\label{ebar}
\end{equation}
The derivation of eq. \ref{ebar} assumes a properly normalized IMF (see eq. \ref{norm} below).
Typical values for $v_\mathrm{esc}^{2}(z)$ at different redshifts can be obtained from  Figure~\ref{fig:velocity}. These result in the following efficiencies for $x=1.3$ and $M_\mathrm{min}=10^{7}\ \mathrm{M_{\odot}}$:
\begin{equation}
\mathit{r}_\mathrm{outflow} \simeq  0.052\ 
\left(\frac{\epsilon}{0.01}\right)
\left(\frac{E_\mathrm{kin}}{10^{51}\ \mathrm{erg}}\right)
\left(\frac{m_\mathrm{inf}}{0.1\ \mathrm{M_\odot}}\right)^{0.3}
\left(\frac{v_\mathrm{esc}^{2}(z)}{3\times 10^{48}\ \mathrm{erg/M_{\odot}}}\right)^{-1}
\end{equation}
for a normal mode of stellar formation at $z\sim 1$,
\begin{equation}
\mathit{r}_\mathrm{outflow} \simeq  1.9\ 
\left(\frac{\epsilon}{0.01}\right)
\left(\frac{E_\mathrm{kin}}{10^{51}\ \mathrm{erg}}\right)
\left(\frac{m_\mathrm{inf}}{0.1\ \mathrm{M_\odot}}\right)^{0.3}
\left(\frac{v_\mathrm{esc}^{2}(z)}{8\times 10^{46}\ \mathrm{erg/M_{\odot}}}\right)^{-1}
\end{equation}
for a normal mode of stellar formation at $z\sim 8$,
\begin{equation}
\mathit{r}_\mathrm{outflow} \simeq  5.8\ 
\left(\frac{\epsilon}{0.01}\right)
\left(\frac{E_\mathrm{kin}}{10^{51}\ \mathrm{erg}}\right)
\left(\frac{m_\mathrm{inf}}{40\ \mathrm{M_\odot}}\right)^{-1}
\left(\frac{v_\mathrm{esc}^{2}(z)}{2\times 10^{46}\ \mathrm{erg/M_{\odot}}}\right)^{-1}
\end{equation}
for a massive mode of stellar formation at $z\sim 16$ ($m_\mathrm{inf}=40\ \mathrm{M_{\odot}}$) and
\begin{equation}
\mathit{r}_\mathrm{outflow} \simeq  8.2\ 
\left(\frac{\epsilon}{0.01}\right)
\left(\frac{E_\mathrm{kin}}{5\times 10^{51}\ \mathrm{erg}}\right)
\left(\frac{m_\mathrm{inf}}{140\ \mathrm{M_\odot}}\right)^{-1}
\left(\frac{v_\mathrm{esc}^{2}(z)}{2\times 10^{46}\ \mathrm{erg/M_{\odot}}}\right)^{-1}
\end{equation}
for a very massive mode of stellar formation at $z\sim 16$ ($m_\mathrm{inf}=140\ \mathrm{M_{\odot}}$). These estimates are confirmed by results described below. 
One should note that the outflow is a self-regulating process and though it may be large
compared with the SFR at a given moment, it will turn off as the SFR decreases. Thus we see that the outflow can be ultra-efficient during the early population III phase and can still be quite efficient at high redshift  even in the normal mode of star formation. In contrast, outflows are very inefficient at  low redshift.

\subsection{Model parameters} 
Before we describe the models in more detail, it will be useful to outline
the adjustable parameters that we will consider. While this list of parameters is not exclusive, 
they are the parameters which most affect the results and shape our conclusions.

\begin{enumerate}

\item The minimum mass $M_\mathrm{min}$ of dark matter halos of star-forming structures.
As described above, the distribution of star-forming structures spans a mass range with
some minimal halo mass.  Structures with masses smaller than $M_\mathrm{min}$ are
considered as part of the IGM and do not participate in the star formation process.
As we will see, when the initial baryon fraction is fixed, 
there is a tight relationship between $M_\mathrm{min}$ and the redshift
at which star formation begins. There are few direct constraints on $M_\mathrm{min}$, and
we will consider values from $10^6$ -- $10^{11}$ M$_\odot$.

\item The initial fraction $f_\mathrm{b,struct}(z_\mathrm{init})$ of baryons within the structures when  star formation begins. We will for the most part assume that star formation begins when the 
baryon fraction in a structure is 1\%. For a given value of $M_\mathrm{min}$, eq. (\ref{eq:fb})
can be used to determine the  redshift
at which star formation begins.  Thus, we can alternatively specify, $z_\mathrm{init}$, and
compute the initial baryon fraction.

\item The efficiency of the outflow $\epsilon$. This parameter too, is largely undetermined.
Its value can be fixed within the model as it sensitively controls the final baryon fraction in structures,
$f_\mathrm{b,struct}(z=0)$.  For models with an exponentially decreasing star formation rate
(as we consider below), increasing $\epsilon$ leads to a decrease in the final value of $f_\mathrm{b,struct}(z=0)$
which we constrain to lie in the range of $61\pm 18\ \%$ \citep[see e.g.][]{fukugita:04} resulting 
in values of $\epsilon < 0.02$.

\item The star formation rate $\Psi(t)$. In principle, there are many possible analytical forms 
for the star formation.  For example, it is common to assume that the SFR is proportional
to the gas mass fraction, $\sigma$, (or some power of $\sigma$). As we describe below, 
we cannot obtain satisfactory fits to both the cosmic SFR and the observed
type II supernova rates when $\Psi \propto \sigma$ for the normal mode  ((0.1 -- 100) M$_\odot$)
of star formation.  In contrast,  we find very good fits with an exponentially decreasing rate
of the form $\Psi = \nu_1 e^{-t/\tau}$. Similarly, we assume an exponential form
for the massive population III mode as well, although here we assume a metallicity-dependent cutoff
so that $\Psi = \nu_2 e^{-Z/Z_\mathrm{crit}}$. The timescale $\tau$ is determined largely by  the 
observed cosmic star formation rate (at low redshift), while the intensity $\nu_1$ is constrained by the observed metal abundances in the ISM and supernova rates.  $\nu_2$ is
 taken to be as large as possible in order to achieve a high ionizing flux at high redshift, while avoiding the overproduction of metals which would appear as a prompt initial enrichment of the ISM and the IGM.
Finally, we assume a critical metallicity of $10^{-4}$ \citep{bromm:03,yoshida:04},
though our results do not change significantly if $Z_\mathrm{crit}$ were a factor of 10 higher.

\item The initial mass function of stars $\Phi(m)$. 
As for the SFR, we must specify an IMF for the normal and massive modes. 
In each case, we take $\Phi(m) \propto m^{-(1+x)}$ for $m_\mathrm{inf}\le m \le m_\mathrm{sup}$, with 
\begin{equation}
\int_{m_\mathrm{inf}}^{m_\mathrm{sup}} dm\ m \Phi(m)=1\ .
\label{norm}
\end{equation}
The normal mode is assumed to contain stars between 0.1 and 100 M$_\odot$.  The slope
of the IMF is constrained by both the type II supernova rate and the metallicity of the ISM.

\item Specific parameters for type Ia supernovae.
 Modelling type Ia supernovae requires several additional parameters which include the mass range $[m_\mathrm{SN\ Ia,min};m_\mathrm{SN\ Ia,max}]$ of the progenitors, the fraction $f_\mathrm{SN\ Ia}$ of these progenitors which will produce a type Ia supernova, and time delay $\Delta t_\mathrm{SN\ Ia}$ between the formation of the white dwarf and the explosion (so that for a progenitor of mass $m$, the total delay between the star formation and the type Ia supernova is $\tau(m)+\Delta t_\mathrm{SN\ Ia}$).

\end{enumerate}


\section{The normal mode of star formation}
\label{sec:model0}

In all of the models considered below, there is a normal mode of star formation 
comprising of stars with masses between 0.1 and 100 M$_\odot$. 
Once the normal mode is adjusted, a massive population III mode is added at high redshift for $Z < Z_\mathrm{c}$. Because the massive mode is cut off rather early (its main effect is to supply a prompt initial enrichment of the ISM and IGM as well as provide an ionizing source at high redshift), most observable constraints are tied to the normal mode.  Thus,  we begin with a more detailed description of the normal mode and its observational consequences.

We listed the main parameters of the model in the previous section. 
The model consists of a superposition of a normal mode of star formation and an early massive mode at high redshift, when the global metallicity in the star-forming structures is still very low (population III stars). We will describe this mode in section~\ref{sec:popIII}. We focus here on the normal mode of star formation. We fix the mass range of star formation to $m_\mathrm{inf}=0.1\ \mathrm{M_{\odot}}$ and $m_\mathrm{sup}=100\ \mathrm{M_\odot}$ so that the only parameter needed to define the IMF is its slope, namely $x_{1}$, which is usually estimated to be in the range $1.30$ \citep{salpeter:55} to $1.7$ \citep{scalo:86}. 

As noted above, we employ an exponentially decreasing SFR (which is representative of elliptical galaxies) as parametrizations such as a Schmidt-law yield significantly poorer fits to the
observational data.  Hence we take, 
$$
\Psi(t) = \nu_{1} \exp{(-(t-t_\mathrm{init})/\tau_{1})}\ ,
$$
where $t_\mathrm{init}$ is the initial age of the Universe at $z_\mathrm{init}$ where the star formation starts in the model, $\tau_{1}$ is a timescale of the order of $2-3\ \mathrm{Gyr}$ and $\nu_{1}=f_{1} m_\mathrm{struct}(t) / \tau_{1}$ with $f_{1}$ being a fraction governing the efficiency of the star formation.\\

Since many of the observational constraints used to fix the parameters pertain to relatively
low redshift, we performed a detailed scan of the parameter space including 
only the normal mode.  The parameter grid was chosen to be:
$M_\mathrm{min}=10^{6}$, $10^{7}$ , $10^{8}$, $10^{9}$, and $10^{11}\ \mathrm{M_{\odot}}$; 
$\epsilon=0$, $10^{-3}$, $2\times 10^{-3}$, $3\times 10^{-3}$, 
{\bf $5\times 10^{-3}$}, $7\times 10^{-3}$, $10^{-2}$, $2\times 10^{-2}$ and $3\times 10^{-2}$; 
$\nu_{1}=0.1$, 0.2, 0.3, 0.4, 0.5, 0.6 and $0.7\ \mathrm{Gyr^{-1}}$; 
$\tau_{1}=2.2$, 2.4, 2.6, 2.8, 3.0, 3.2 Gyr; $x_1=1.3$, 1.35 and 1.4. 
In all cases, we have assumed that the onset of star formation begins when the baryon fraction 
in structures is 1\%. We have tested that this initial fraction has a very weak impact on the results concerning the normal mode of star formation. It is on the other hand of great importance for the population III stars and its effect will be studied in section~\ref{sec:popIII}. One should also note that there are additional parameters associated with the rate of type Ia supernovae. These are directly constrained by observations and will be fixed in section~\ref{sec:SNae}.\\ 

To determine the parameters which best fit the observations, we performed $\chi^2$ analysis
over the parameter grid. Included in the $\chi^2$ analysis are six sets of observational data:
(1) The observed cosmic star formation rate up to $z\sim 5$ \citep{hopkins:04}. The data were
binned and averaged in redshift leading to somewhat larger observational uncertainties
at a given redshift than typically reported for a single measurement;
(2) The observed rate of type II supernovae up to $z\sim 0.7$ \citep{dahlen:04};
(3) The present fraction of baryons in structures, $f_\mathrm{b,struct}(z=0)\simeq 61\pm 18\ \%$
\citep{fukugita:04}; 
(4) The present fraction of baryons in stars, $f_\mathrm{b,*}(z=0)\simeq 6\pm 6\ \%$
\citep{fukugita:04}; 
(5,6) The evolution of the metal content in the ISM and IGM from $z\sim 5$ to $z=0$. 
Specifically, for the ISM we took   $\log_{10}{Z/Z_\mathrm{\odot}}=-0.3\pm 0.2$ at $z=0$ and $-0.9\pm 0.2$ at $z=4$. 
For the IGM we took, 
 $\log_{10}{Z/Z_\mathrm{\odot}}=-2.5\pm 0.2$ at $z=0$ and $-3.1\pm 0.2$ at $z=4$ 
\citep{prochaska:03,ledoux:03,songaila:01,schaye:03,aguirre:04}. These constraints are chosen so that the metallicity in the ISM (IGM) is always larger (smaller)  than in the most (least) metallic DLA.
The dominant effect
of including the massive mode will fall on the metallicity of the IGM and will be discussed in section 5 below.

The results of the $\chi^2$ minimization are listed in Table~\ref{tab:model0} separately for each
value of $M_\mathrm{min}$.
Having fixed the onset of star formation to correspond to an initial baryon fraction of 1\%, each
value of $M_\mathrm{min}$ leads to a distinct redshift for initial star formation.  This value of $z_\mathrm{init}$ is also given in the table.
In each case, we found a best fit IMF slope of $x_1 = 1.3$.
It should be remembered that when we add in the massive mode, these `best' fits
have to be modified in order to take into account the metals produced by population III stars.
This will affect only the efficiency of the outflow, $\epsilon$, as the other parameters
control the later phases of evolution.
Of the sampled values of $M_\mathrm{min}$, $M_\mathrm{min} = 10^7$ M$_\odot$ was
found to give the best fit.
However if the observed SFR at high redshift ($z>3$) has been underestimated, 
the best fit model may be at lower values of $M_\mathrm{min}$.  For example, we find that  $M_\mathrm{min}=10^{6}\ \mathrm{M_{\odot}}$ becomes a better fit if the high redshift SFR data are increased by a factor $\sim 3$. 
Also shown in the table are the output values for the baryon fraction in structures and in stars.
Note that none of the models accurately reproduce the stellar baryon fraction at $z = 0$.
In each case, we over-produce stars at a level of about 1$\sigma$.

\begin{table}[t]
\begin{center}
\begin{tabular}{ccccccc}
\multicolumn{7}{c}{\textbf{Normal mode}}  \\ 
$M_{\mathrm{min}}$ & 
$z_{\mathrm{init}}$ & 
$\epsilon$ & 
$\nu_{1}$  & 
$\tau_{1}$ &
$f_{\mathrm{b,struct}}$ &
$f_{\mathrm{b,*}}$ \\
($M_{\odot}$) & 
 & 
 & 
(Gyr$^{-1}$) & 
(Gyr) &
\multicolumn{2}{c}{$(z=0)$} \\
\hline
\hline
$10^{6}$  & 18.2 & $2\times 10^{-3}$ & 0.2 & 2.8 & 63 \% & 14 \%  \\ 
\hline
$\mathbf{10^{7}}$  & \textbf{16.0} & $\mathbf{3\times 10^{-3}}$ & \textbf{0.2} & \textbf{2.8} & \textbf{61 \%} & \textbf{13 \%} \\ 
\hline
$10^{8}$  & 13.7 & $5\times 10^{-3}$ & 0.2 & 2.8 & 58 \% & 11 \% \\ 
\hline
$10^{9}$  & 11.3 & $        10^{-2}$ & 0.2 & 3.0 & 54 \% & 11 \%  \\ 
\hline
$10^{11}$ & 6.57 & $1.5\times 10^{-2}$ & 0.5 & 2.2 & 44 \% & 11 \% \\ 
\end{tabular}
\end{center}
\caption{The best fit models for the normal mode of star formation (Model 0). Column 1 indicates the input value of the minimum mass for star forming structures. Column 2 is derived from column 1, having assumed that $f_b = 1\%$ when star formation begins.  In columns 3, 4, and 5, we show the results of the minimization for each value of $M_\mathrm{min}$.  The slope of the IMF is $x=1.3$ for all these best models. 
Among all these models, the best fit is obtained for $M_\mathrm{min}=10^{7}\ \mathrm{M_{\odot}}$ (bold face). For this reason, this model is used as a reference in the figures. 
Columns 6 and 7 contain the output values of the baryon fraction in structures and in stars respectively. 
}
\label{tab:model0}
\end{table}

\begin{figure}
\begin{center}
\resizebox{\textwidth}{!}{\includegraphics{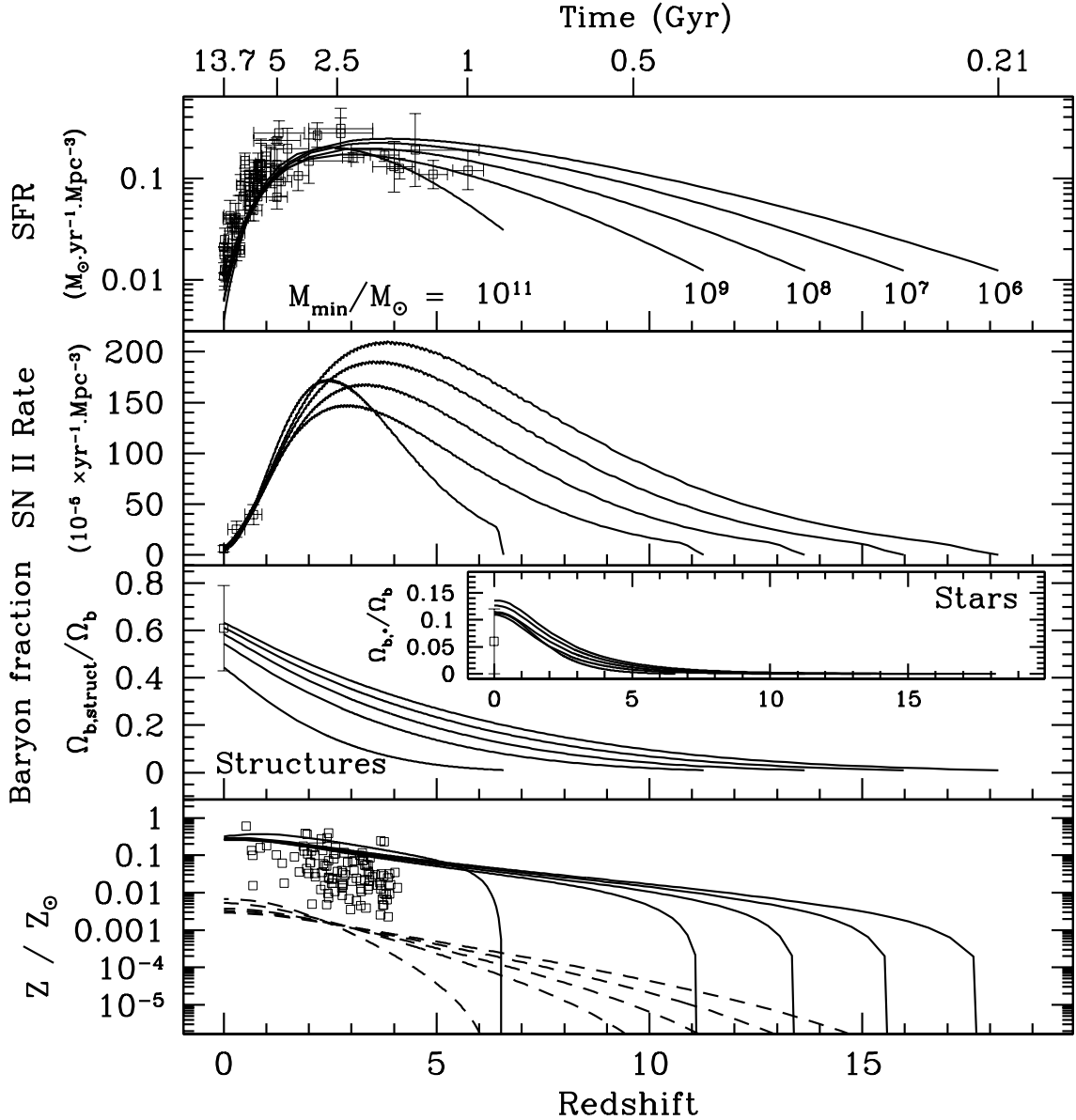}}
\end{center}
\caption{The cosmic star formation rate, type II supernova rate, baryon evolution and global metallicity for the normal mode of star formation (Model 0). Plotted are the results of our best fit models for each choice of $M_\mathrm{min}$. In the top panel, we show the SFR for each model.
The observed SFR up to $z\sim 5$ is taken from \citet{hopkins:04}. In the second panel, we show the predicted rate for type II supernovae. The observed rates up to $z \sim 0.7$ is taken from \citet{dahlen:04}. In the third panel we show the evolution of the
baryon fraction in structures and in stars (inset). 
The local values at $z=0$ are taken from \citet{fukugita:04}.
In the bottom panel, we show the evolution of the metallicity in both the ISM and IGM (dashed curves). The metallicity of 100 Damped Lyman-$\alpha$ systems as measured by \citet{prochaska:03} is also plotted. }
\label{fig:model0SFRZ}
\end{figure}

The results of our best fit models are plotted in Figure~\ref{fig:model0SFRZ} for 
each value $M_\mathrm{min}$ as indicated in the upper panel. Because we have fixed the initial baryon fraction in structures, each value of $M_\mathrm{min}$ corresponds to a different initial redshift. 
As one can see, each of the models gives a satisfactory fit to the global SFR, save perhaps the case
with $M_\mathrm{min} = 10^{11}$ M$_\odot$. The data shown (taken from \citet{hopkins:04} 
has already been corrected for extinction. 
If we compare the cosmic SFR calculated by  \citet{croton:05a} (see also \citet{springel:03}), we find good agreement at high redshift, particularly with the  $M_\mathrm{min} = 10^{7}$ M$_\odot$ model. However, at $z<5$ their SFR is somewhat lower than ours by a factor of about 2.  Moreover, in our model, about 60 per cent of all stars formed by z = 3 compared with 
50 per cent at the same redshift in their model.
The predicted type II supernova rates are also an excellent fit to the 
existing data at low redshift. The predicted rates are substantially higher at higher redshift and lower
$M_\mathrm{min}$. 

\begin{figure}
\begin{center}
\resizebox{\textwidth}{!}{\includegraphics{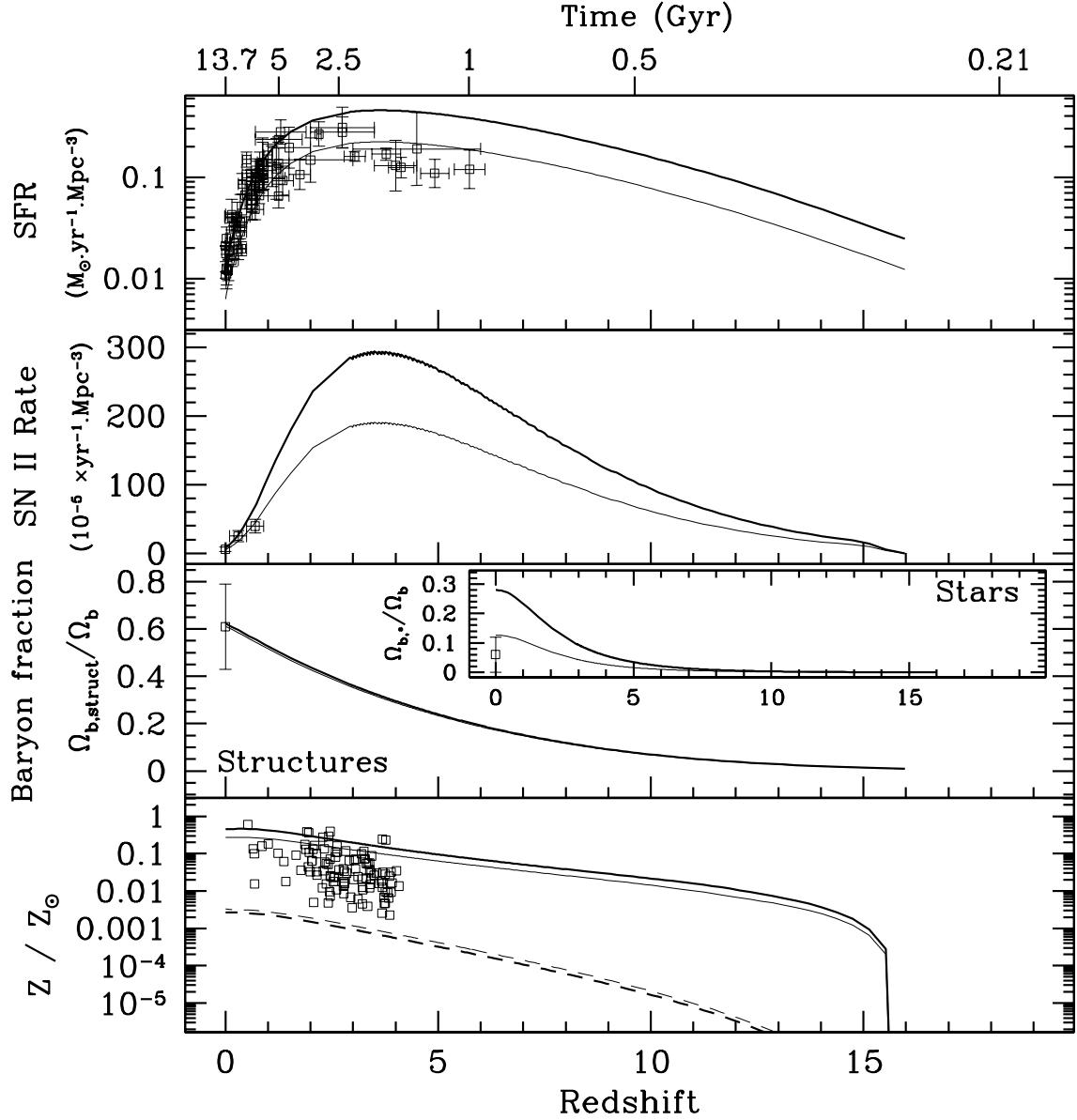}}
\end{center}
\caption{As in  Figure~\ref{fig:model0SFRZ}, we show the best fit model with $M_\mathrm{min}=10^{7}\ \mathrm{M_{\odot}}$ (thin line) and the same model where the astration rate $\nu_{1}$ has been arbitrarily multiplied by a factor 2 (thick line). Note the new high value of the corresponding local fraction of baryons in stars and the increase in the peak value of the SN rate at a redshift of about 3.}
\label{fig:model0HighSFR}
\end{figure}

As noted above, while the baryon fraction in structures is fit quite well, 
the baryon fraction in stars  at z = 0 (or equivalently $\Omega_b^*$) is somewhat high. This output
is clearly linked to the SFR which is in fact well fit and could be even higher than shown here.
For example, we show in Figure~\ref{fig:model0HighSFR} resulting predictions 
when the astration rate $\nu_1$ has been doubled for our best fit model with 
$M_\mathrm{min}=10^{7}\ \mathrm{M_{\odot}}$.  As one can see, the baryon fraction in stars
now rises to nearly 30\% and clearly disagrees with observational determinations of this quantity.
Finally, coming back to Figure~\ref{fig:model0SFRZ}, we also show the evolution of the metallicity in both the ISM and IGM (dashed curves). 
While the ISM metallicity rises very quickly initially,  evolution is more gradual in the IGM
where it is controlled by the outflow. 
Furthermore, we see that the evolution of the metallicity in the IGM is dependent on
$M_\mathrm{min}$. Since models with large $M_\mathrm{min}$ begin rather late (e.g.
the model with $M_\mathrm{min} = 10^{11}\ \mathrm{M_{\odot}}$ begins at $z \approx 6.5$)
these models have difficulty in producing a suitably large metallicity in the IGM at redshift 2 - 3.
Note that we have only included the normal mode here,
and contributions from population III stars will only serve to increase the metallicity in both the ISM and IGM.

\section{Supernova rates}
\label{sec:SNae}

Among the important constraints imposed on these models of cosmic chemical evolution
are the rates of type Ia and II supernovae. These are intimately linked to the
choice of an IMF and SFR and represent an independent probe of the
star forming universe at high redshift. While type II rates directly trace the star formation history,
there is a model dependent time delay between the formation (and lifetime) of the 
progenitor star and the type Ia explosion.  Since existing data is available only for relatively low
redshifts, we discuss the predicted SN rates in the context of Model 0 (the normal mode of star
formation). Indeed, at these redshifts the massive mode is not operational.

\subsection{Type II supernovae}

\begin{figure}
\begin{center}
\resizebox{\textwidth}{!}{\includegraphics{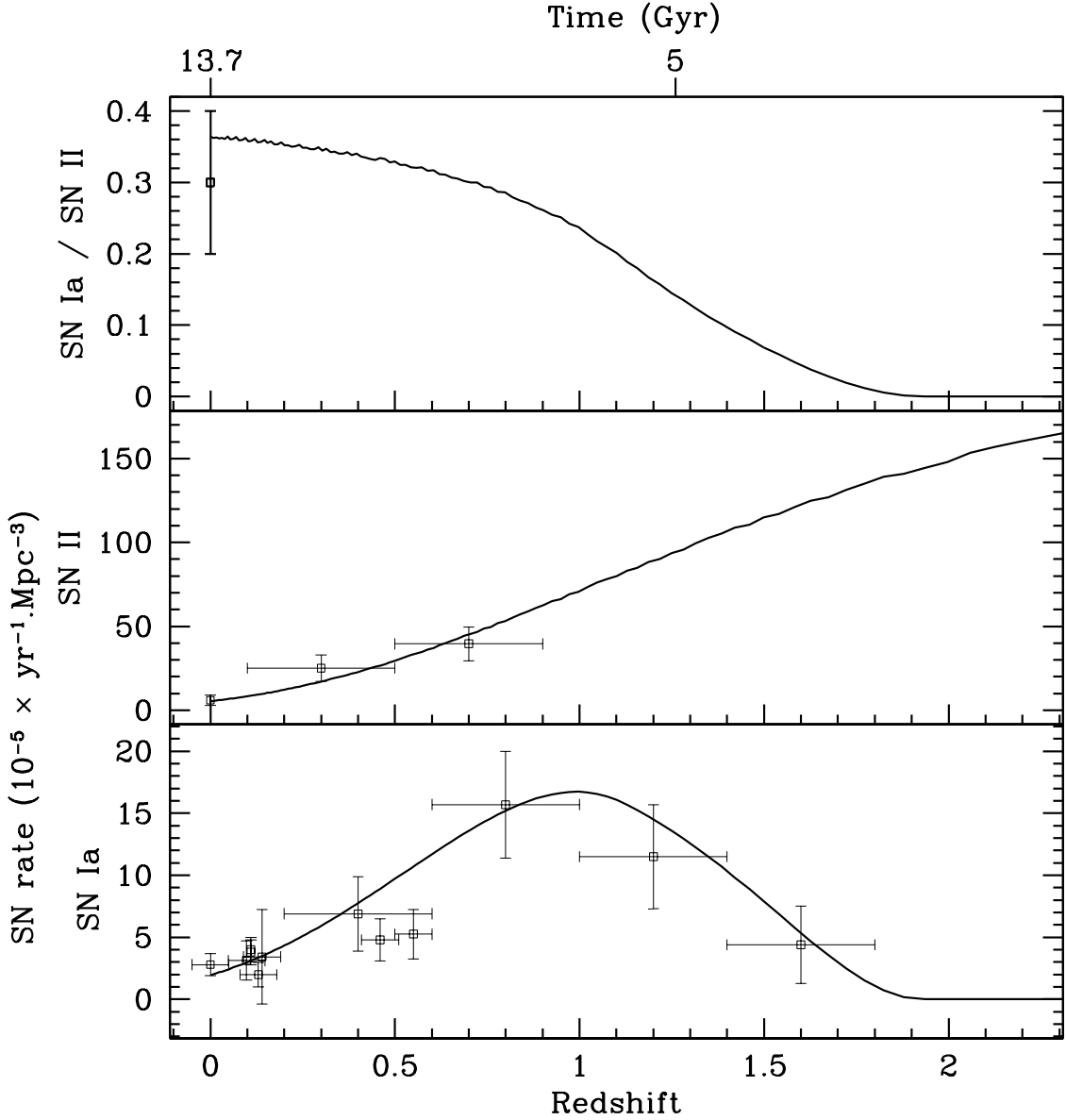}}
\end{center}
\caption{The evolution of the ratio of type Ia to type II supernova event rates is shown for the best fit model with $M_\mathrm{min}=10^{7}\ \mathrm{M_{\odot}}$ in the upper panel. The observed local values are taken from \citet{cappellaro:99}. The type II SN rate as a function of redshift is shown in the middle panel. The observed rates at $z\sim 0.3$ and $z \sim 0.7$ are taken from \citet{dahlen:04}. In the bottom panel, we show the  type Ia SN rate as a function of redshift. The observed rate up to $z\sim 1.6$ is taken from \citet{reiss:00}, \citet{hardin:00}, \citet{pain:02}, \citet{madgwick:03}, \citet{tonry:03}, \citet{strolger:04}, \citet{blanc:04} and \citet{dahlen:04}.}
\label{fig:model0SN}
\end{figure}

The predicted rates for type II supernovae were discussed briefly in the previous section,
where we showed the results as a function of redshift for several choices of $M_\mathrm{min}$ in
Figure~\ref{fig:model0SFRZ}. In Figure~\ref{fig:model0SN}, we focus on the low redshift range
and concentrate on the best fit model with $M_\mathrm{min}=10^{7}\ \mathrm{M_{\odot}}$.
The data shown are taken from the Great Observatories Origins Deep Survey (GOODS) \citep[][and references therein]{dahlen:04}. Of the 42 supernovae observed, 25 are associated with type Ia
and 17 with core collapse supernovae (type II as well as type Ib,c). 

The GOODS data for core collapse supernovae have been placed in two bins at $z = 0.3 \pm 0.2$ and 
$z = 0.7 \pm 0.2$.  Their results (which have been corrected for the effects of extinction) show SN rates which are significantly higher than the local rate (at $z = 0$) which is taken from \citet{cappellaro:99}. Since the timescale between star formation
and the core collapse explosion is very short (particularly for very massive stars), there will be
no contribution of Population III stars to type II rates at $z \le 1$.  

\begin{figure}
\begin{center}
\resizebox{\textwidth}{!}{\includegraphics{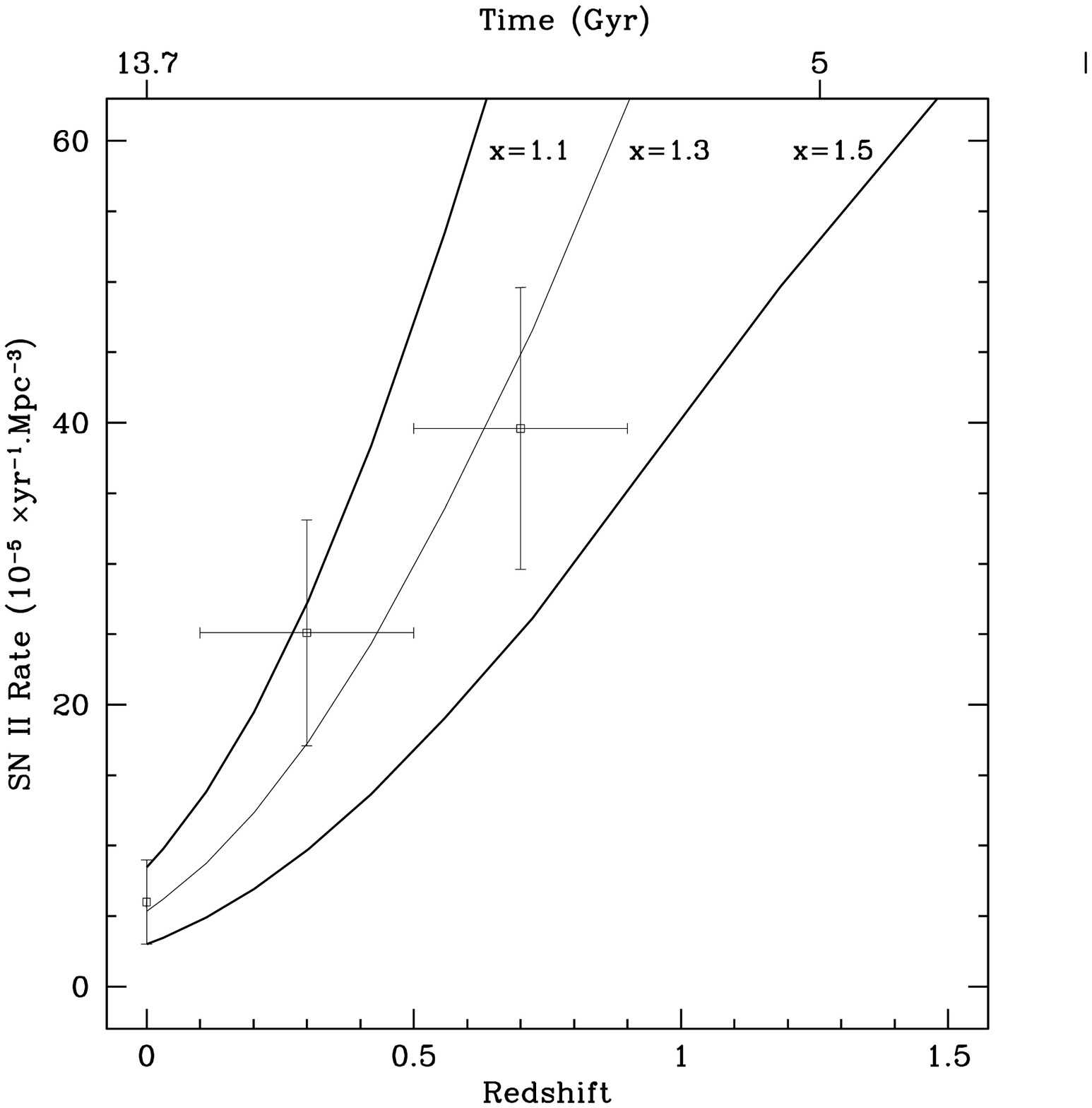}}
\end{center}
\caption{The effect of the IMF slope on the SNII rate as a function of redshift. The reference model (plotted as the thin curve) assumes 
$M_\mathrm{min} = 10^{7}\ \mathrm{M_{\odot}}$ and an IMF slope of $x_1 =1.3$). For comparison, two other models with slopes $x=1.1$ and $x=1.5$ are also plotted. The data shown are the same as in Figure~\ref{fig:model0SN}.}
\label{fig:SNII}
\end{figure}

As remarked above, the SN type II rate is directly related to the overall SFR and therefore
to the assumed astration rate $\nu_1$ and our best fit model
does an excellent job at describing the existing data. 
Note that, had we chosen a Schmidt law ($\Psi \propto \sigma_{\mathrm gas})$,
it would not be possible to simultaneously fit both the local ($z = 0$) data as well as the GOODS data.
In general, models with a Schmidt law can not account for a large enhancement in the 
SN rate. Therefore once the peak of the SFR has been fixed (at $z \approx 1$),
the local SN rate is too large.
The slope of the type II rate
(with respect to redshift) is also dependent on the IMF slope, $x_1$.  In Figure~\ref{fig:SNII},
we show this dependence displaying the predicted type II rates for $x_1 = 1.1, 1.3$ and 1.5.
Of course the choice of $\nu_1$ and $x_1$ also has a strong effect on the metallicity and
detailed chemical history as discussed in more detail in the next section when Population III
stars are included in the analysis.

It is important to note that while the data at $z \le 1$ is well described by the model, the same model
predicts much higher SN rates at higher redshift.  Indeed, the SN type II rate peaks at
a redshift $z \approx 3$ at a rate which is nearly 8 times the observed rate at $z \approx 0.7$. 
It will be interesting to see whether future data will be able to probe the SN rate at higher redshift.
These elevated rates may be detectable in spite of the expected increase in dust extinction due to the
early production of metals.  An alternative to the direct detection of SN at high redshift is the indirect detection of SN through neutrinos.  The predicted neutrino background is very sensitive to the
assumed chemical evolution model \citep{ando:04,iocco:05,dosv:05}.

\subsection{Type Ia supernovae}

As in the case for type II supernovae, the data (also taken from GOODS \citep{dahlen:04}) is binned
into four redshift bins at $z = 0.4$, 0.8, 1.2, and 1.6, each with a spread of $\pm 0.2$.
These are shown in the lower panel of Figure~\ref{fig:model0SN} along with data from  \citet{reiss:00}, \citet{hardin:00}, \citet{pain:02}, \citet{madgwick:03}, \citet{tonry:03}, \citet{strolger:04}, and \citet{blanc:04}. Once again, the local ($z = 0$) data is taken from \citet{cappellaro:99}. Unlike the type II data,
there appears to be a definite peak in the rate of type Ia supernovae around $z \sim 1$. 

As the progenitors of  type Ia supernovae are stars with  intermediate masses between 
2 and 8 M$_\odot$,
the rate for SN type Ia is controlled entirely by parameters within the normal mode of star formation.
There are however two additional parameters that must be adjusted to fit the observed data, 
namely the fraction of intermediate mass stars that explode as SN type Ia, and the time delay between the death of the progenitor and the SN explosion.  In most previous 
models of Galactic chemical evolution, this delay was taken to be of order 1 Gyr. Recently, many studies have been devoted to the progenitors of SNIa supernovae, including the analysis of time delays and the mass ranges (\citet{han:04}, \citet{yungelson:04}, \citet{garnavich:05} and \citet{belczynski:05}). The GOODS data are now consistent with significantly larger delays (of order 4 Gyr, including the lifetime of the star)
and exclude delays less than 2 Gyr at the 95~\% confidence level \citep{strolger:04,dahlen:04}.
In the context of the normal mode (Model 0),
we find that these observations are well reproduced if 
$\sim 1\ \%$ of intermediate mass stars lead to type Ia supernovae and if the typical delay between the formation of the white dwarf and the explosion is $\sim 3-3.5\ \mathrm{Gyr}$.
The resulting comparison between our theoretical prediction and the data is shown in the lower
panel of Figure~\ref{fig:model0SN} where the effect of the time delay is quite apparent and is 
chiefly responsible for the peak in the SN rate at $z \sim 1$. While this comparison is made using our best fit model with $M_\mathrm{min} = 10^{7}\ \mathrm{M_{\odot}}$, the resulting rate for type Ia supernovae would be quite similar for other choices of $M_\mathrm{min}$.
For an alternative approach see \citet{scann:05,mannucci:05}.

\begin{figure}
\begin{center}
\resizebox{\textwidth}{!}{\includegraphics{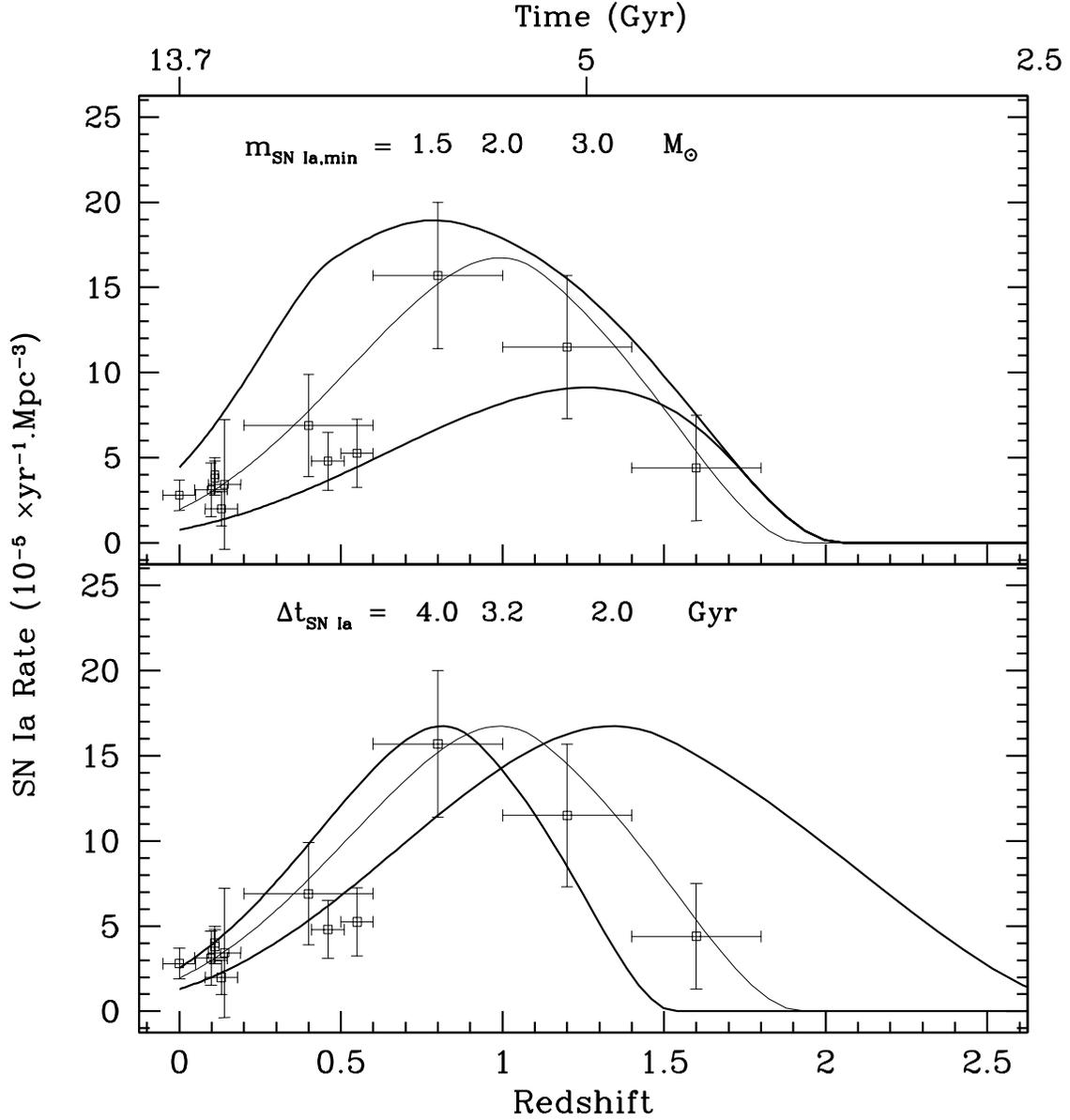}}
\end{center}
\caption{The effect of the progenitor mass range and of the time delay on the SNIa rate.
The thin curves correspond to the reference model with  $M_\mathrm{min} = 10^{7}\ \mathrm{M_{\odot}}$, a minimal progenitor mass of 2 M$_\odot$, and a time delay of 3.2 Gyr. The sensitivity to the lower end of the intermediate mass range is shown in the upper panel where two other models with a minimum mass of 1.5 and 3 $\mathrm{M_{\odot}}$ are also indicated. The sensitivity to the time delay 
is shown in the lower panel where
two other models with time delays of 2 and 4 Gyr are also indicated. The data shown are the same as in Figure~\ref{fig:model0SN}.}
\label{fig:SNIa}
\end{figure}

To test the sensitivity of the result to the time delay, we show in Figure~\ref{fig:SNIa}
the predicted SNIa rates using a time delay of 2.0 and 4.0 Gyr, compared with our 
best fit value of 3.2 Gyr.  Also shown in Figure~\ref{fig:SNIa} (upper panel) is the sensitivity to the
adopted intermediate mass range.  Our best fit model assumes a range of 2 - 8 M$_\odot$.
Also plotted are models where the lower end of the mass range is 1.5 and 3 M$_\odot$.

Finally, in the upper panel of Figure~\ref{fig:model0SN}, we compare our predictions of type Ia to type II
supernovae.  We see that these are  quite consistent with the local rate \citep{cappellaro:99}.
The ratio drops off at redshifts $z \ga 1$ due to the  time delay in producing type Ia events.
Furthermore, when SN Ia explosions do occur, structures are larger and mass outflows are
suppressed (see discussion in the next section). 
Thus, as a consequence of the SN Ia time delay, we predict that while $\sim 50\ \%$ of iron in structures is produced by type Ia supernovae, this fraction is only $\sim 10\ \%$ in the IGM.


\section{Population III stars}
\label{sec:popIII}

\subsection{Necessity of an initial starburst of massive stars}

As shown in \citet{daigne:04}, the normal mode of star formation, labelled Model 0 there as well as here,
is not capable of reionizing the Universe at high redshift.  Indeed, at a redshift $z = 17$, 
only 1.6 ionizing photons per baryon are available compared to the requisite value of approximately 20
assuming a clumpiness factor, $C_\mathrm{H\;\scriptscriptstyle{II}} = 10$ \citep{ricotti:04b}.  
It was also shown in 
\citet{daigne:04} that Model 0 alone is not capable of explaining the observed abundance
patterns in the extremely iron-poor stars, CS 22949-037 \citep{depagne:02,israelian:04},
 HE 0107-5240 \citep{christlieb:04,bessell:04}, HE 1327-2326  (\citet{frebel:05}) and G 77-61 (\citet{plez:05}).

The necessity of a massive mode of star formation, active at high redshift has become an integral part of
our emerging picture of the growth of galactic structures. 
However, there is an active debate as to the exact nature of the massive mode.
As in \citet{daigne:04}, we consider three possibilities for the massive mode.
1) stars with masses in the range 40 -- 100 M$_\odot$, Model 1; these stars terminate
as type II supernovae.
2) stars with masses in the range 140 -- 260 M$_\odot$, Model 2a; these stars terminate
as pair-instability supernovae (PISN).
3) stars with masses in the range 270 -- 500 M$_\odot$, Model 2b; these stars terminate
as black holes through total collapse and do not contribute any metal enrichment.

In all cases, we assume a bimodal birthrate function of the form
\begin{equation}
B(m,t,Z) = \Phi_1(m) \Psi_1(t) + \Phi_2(m) \Psi_2(Z)
\end{equation}
where the two IMFs are normalized independently as in eq. (\ref{norm}).
The SFR, $\Psi_2$, is now expressed as a function of metallicity and cuts off
once a critical metallicity is reached, as described below.
Furthermore, we assume the same slope of the IMF, $x_1 = x_2 = 1.3$ so that 
$\Phi_1$ and $\Phi_2$ differ only in their mass range. In principle, we can choose to start 
the normal mode of star formation either simultaneously with the massive mode,
or sequentially, that is, when the $Z > Z_\mathrm{crit}$. In the latter case, one
can argue that the initial injection of metals by Pop III stars is responsible for the 
formation of the first extremely metal poor Pop II (or Pop II.5) stars \citep{mackey:03,salvaterra:04,johnson:05}.  Since critical metallicity is achieved very rapidly (within 3 Myr), our results
for these two choices are almost indistinguishable.

As the cooling process of the gas depends strongly on its chemical composition, 
it is believed that the evolution of the mass range of the IMF is mainly governed by the global metallicity
\citep{fang:04}. As noted above, we assume a transition from  population III to the normal formation mode at a critical metallicity $Z_\mathrm{crit}$ \citep{bromm:03,yoshida:04} by defining the SFR of the massive mode by
\begin{equation}
\Psi(t)_2 = \nu_{2} \exp{\left(- Z / Z_\mathrm{crit}\right)}\ ,
\end{equation}
with $\nu_{2} = f_{2} m_\mathrm{struct}(t)$. We adopt $Z_\mathrm{crit}/Z_\mathrm{\odot}=10^{-4}$ .

Because the massive stars adopted in Models 1 and 2a are efficient in producing heavy elements,
the duration of the massive phase in these models is relatively brief.  In contrast, no heavy elements are produced in Model 2b, and therefore, the massive mode continues to affect the evolution of structures
until the metallicity of the {\it normal} mode reaches the critical value, which is also relatively fast, due
to the short lifetimes of massive stars.

Most of the constraints discussed above must be applied at relatively low redshift ($z \la 5$).
As a result they fix the parameters of Model 0.  For the most part these are unchanged, with one
important exception.  The metallicity of the IGM now receives two distinct contributions.
The first from the massive mode, which will appear as a prompt initial enrichment, and the second
from the outflows of the normal mode which contribute at lower redshifts.  In principle, these contributions can be distinguished with precise abundance data as a function of redshift.
If the metallicity in the IGM originates from the massive mode, we expect IGM abundances which are 
constant with respect to $z$.  On the other hand, if there is a significant contribution from the outflows 
due to the normal mode, we should see abundances which vary with redshift.
Unfortunately, the current data does not allow us to distinguish between these two possibilities.
Indeed, one would expect that the IGM contains contributions from both modes,
but the relative contribution in fact becomes another parameter of our model.

In what follows, we will consider two possibilities for the role of the massive mode
in the enrichment of the IGM.  In one case, we will assume that nearly all ($\sim 90 \%$) of
the IGM metallicity originates from massive stars.
Increasing the contribution of the massive mode allows us to maximize its ionizing efficiency.
In section 3, we discussed the role of the normal mode.  There, we had effectively set
$\nu_2 = 0$, and the efficiency of outflow, $\epsilon$, controlled both the
baryon fraction in structures and the metallicity of the IGM.  When $\nu_2$ is maximized, 
we will clearly be forced to reduce the value of $\epsilon$ to avoid the overproduction of metals
in the IGM.  
Despite the low value of $\epsilon$, outflow is more efficient at early times and the enrichment of the IGM is also more efficient due to the high level of SNe (large $\nu_2$). Consequently, in this case,
most of the IGM metals comes from pop III stars and we expect that $Z_\mathrm{IGM}$ will be constant with $z$. Note that although the value of $Z_\mathrm{crit}$ is very small, ISM metallicities and
even IGM metallicities much greater than $Z_\mathrm{crit}$ can be produced by the massive mode.
This is because of the finite lifetimes of the massive stars.  Although these lifetimes are short,
they are nevertheless of order a few Myr and hence metals continue to be injected into the
ISM (and IGM) after $Z > Z_\mathrm{crit}$ is reached and the formation of new massive stars 
is quenched.

We will also consider an intermediate case where the IGM receives 
equal contributions from the massive and normal modes at $z = 2.5$. 
In this case, $\nu_2$ is again chosen to maximize the
efficiency of reionization, and $\epsilon$ will be closer to the value found in section 3.

In Table 2, we display the four parameters for Models 1, 2a, and 2b, for both sets of assumptions
concerning the contribution of Pop III to the IGM. Models 1, 2a, and 2b refer to the case
where the IGM receives equal contributions from the normal and massive modes (at $z \approx 2.5$),
whereas Models, 1e and 2ae, refer to the extreme case where 90\% of the metallicity at
$z \simeq 2.5$ is derived from the massive mode. Note that there is no Model 2be, as
no metals are ejected into either the ISM or IGM from massive mode stars in this case,
and all of the low redshift metallicity is due to the normal mode.
As noted above, the outflow efficiencies for Model 1 are within a factor of 2 
of those for Model 0 (cf. Table 1).  In contrast, the efficiencies for Model 1e are significantly lower.
However as one can see in Table 2, the low outflow efficiency is compensated for by
a large SFR ($\nu_2$) greatly increasing the models capacity for reionization.
The final baryon fraction in structures and stars is very similar in these models
to that found in Model 0. In general, $f_{\mathrm{b,struct}}$ is larger than the values in 
Table 1 by 1 -- 2 \% for Models 1, 2a, and 2b, and by 2 -- 4 \% for Models 1e and 2ae. 
$f_{\mathrm{b,*}}$ is unchanged from Model 0. As we will see, there is also a close correlation between
the intensity of the SFR associated with the massive mode, reionization and the metal enrichment provided by these Pop III stars.

\begin{table}[t]
\begin{center}
\begin{tabular}{cccc}
\multicolumn{4}{c}{\textbf{Massive mode}}  \\ 
Model & $M_{\mathrm{min}}$ & 
$\epsilon$ & 
$\nu_{2}$  \\
& ($M_{\odot}$) 
 & 
 & 
(Gyr$^{-1}$)  \\
\hline
\hline
1 & $10^{6}$  &  $1.5 \times 10^{-3}$ & 30 \\
\hline
1e & $10^{6}$  &  $5 \times 10^{-5}$ & 260 \\
\hline
1 & $10^{7}$  &  $2 \times 10^{-3}$ & 60 \\
\hline
1e & $10^{7}$  &  $6 \times 10^{-5}$ & 340 \\
\hline
1 & $10^{8}$  &  $2.5\times 10^{-3}$ & 80 \\
\hline
1e & $10^{8}$  &  $1.8\times 10^{-4}$ & 290 \\
\hline
1& $10^{9}$  & $    5 \times    10^{-3}$ & 100 \\
\hline
1 & $10^{11}$ &  $7\times 10^{-3}$ & 200 \\
\hline
2a & $10^{7}$  &  $1 \times 10^{-3}$ & 9 \\
\hline
2ae & $10^{7}$  &  $8 \times 10^{-5}$ & 40 \\
\hline
2b & $10^{7}$  &  $3 \times 10^{-3}$ & 100 \\
\hline
\end{tabular}
\end{center}
\caption{Parameter values for the massive starburst Models 1, 2a and 2b. Column 1 indicates the 
model number and column 2 the input value of the minimum mass for star forming structures.   In columns 3 and 4, we show the outflow efficiency and massive mode SFR.  }
\label{tab:model1}
\end{table}

Once the ratio of massive/normal mode contribution to the IGM is adopted,
we apply the observational constraints to fix the values of $\epsilon$ and $\nu_2$.
In addition to the constraint based on the overall metallicity of the IGM, we 
must require that the massive mode produces a sufficient number of ionizing photons. 
We will also look at the evolutionary history of several individual element abundances
including C, O, Si, and Fe.  Finally, we compare the results to the observed abundances of
several extremely iron-poor stars CS 22949-037 \citep{depagne:02,israelian:04}, HE 0107-5240 \citep{christlieb:04,bessell:04}, HE1327-2326 \citep{frebel:05}, and G 77-61 \citep{plez:05}. They represent the mean abundance in the Universe at this epoch.

\subsection{Model 1}

\begin{figure}
\begin{center}
\resizebox{\textwidth}{!}{\includegraphics{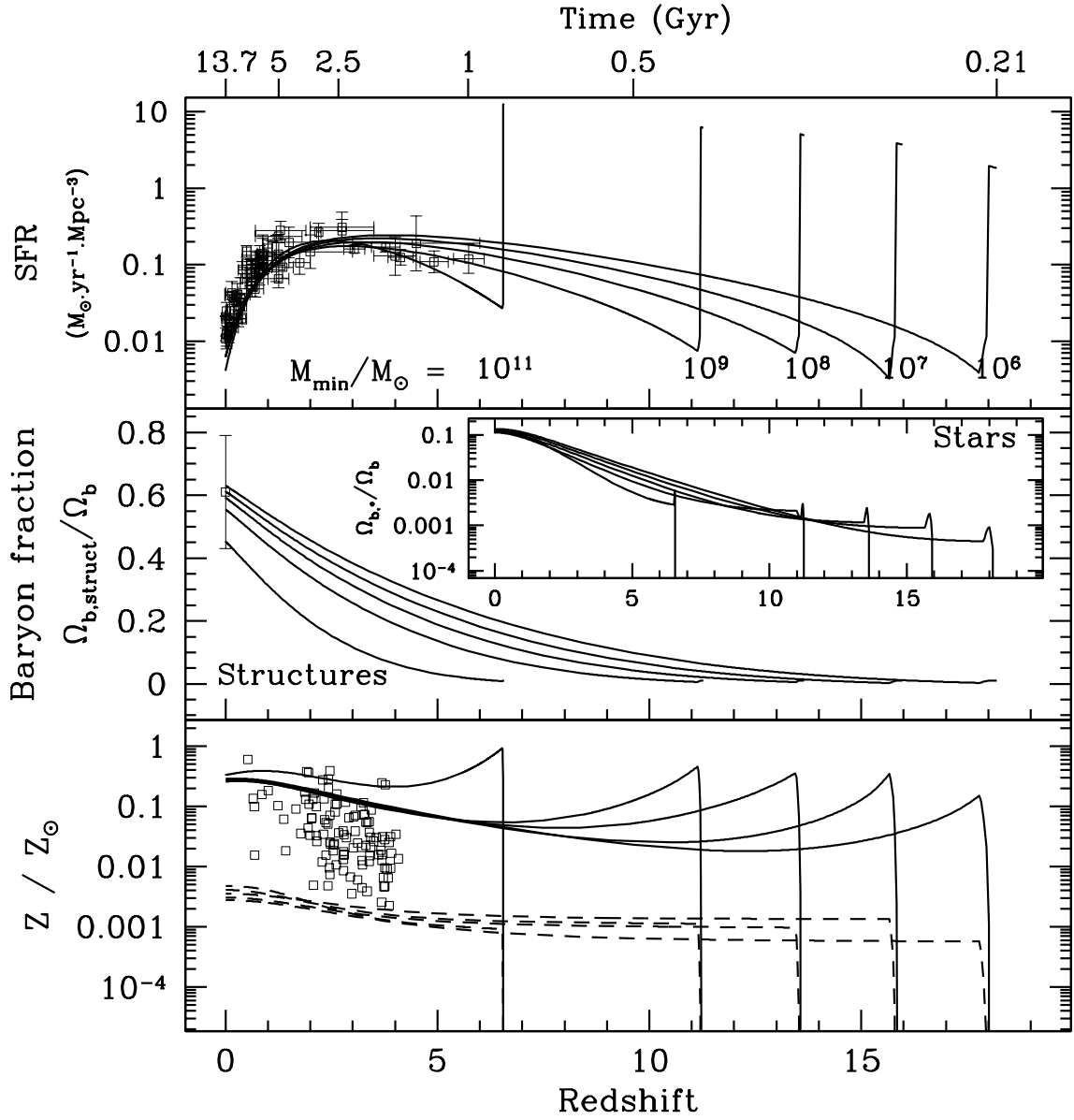}}
\end{center}
\caption{As in Figure~\ref{fig:model0SFRZ} but with an added initial starburst with stellar mass in the range  40--100 $\mathrm{M_{\odot}}$ (Model 1). See Table~\ref{tab:model1} for the parameters used.}
\label{fig:model1SFRstarsZ}
\end{figure}

We begin by discussing the results of the bimodal model which combines the normal mode (Model 0)
with the massive mode denoted as Model 1 which includes stars with masses in the range
40 --100 M $_\odot$.  In this model (as well as in Model 2a described below), star formation 
begins at a very high rate and falls precipitously as metals are injected into the ISM. 
In inhomogeneous models of structure formation, we would expect this narrow burst to broaden \citep{scannapieco:03}. As in the case of Model 0, the onset of star formation is determined by $M_\mathrm{min}$, and the initial value for the baryon fraction in structures (fixed to be 1 \%). Results for the SFR, baryon fraction  and metallicity are shown in Figure~\ref{fig:model1SFRstarsZ} for the intermediate case where
Model 1 stars contribute 50\% to the IGM metallicity at z = 2 to 3 and 20\% at z = 0.
With exception of the metallicity, shown in the third panel, the effect of the Population III stars is
minor as the SFR and baryon fraction is very similar here to the case of Model 0 shown in Figure~\ref{fig:model0SFRZ}.
Note that at high redshift, the fraction of the baryons in massive stars is about $10^{-3}$ as seen in the insert to the middle panel of Figure~\ref{fig:model1SFRstarsZ}.
As one can see in the lower panel, the metallicity in the ISM reaches 
values far in excess of $Z_\mathrm{crit}$ due to the finite lifetime of the massive
stars relative to the speed at which the metallicity is attained. Notice also that
once the metallicity from Pop III stars is produced, the ISM metallicity later decreases as a result of
the accretion of metal-free gas as the structures grow.

\begin{figure}
\begin{center}
\resizebox{\textwidth}{!}{\includegraphics{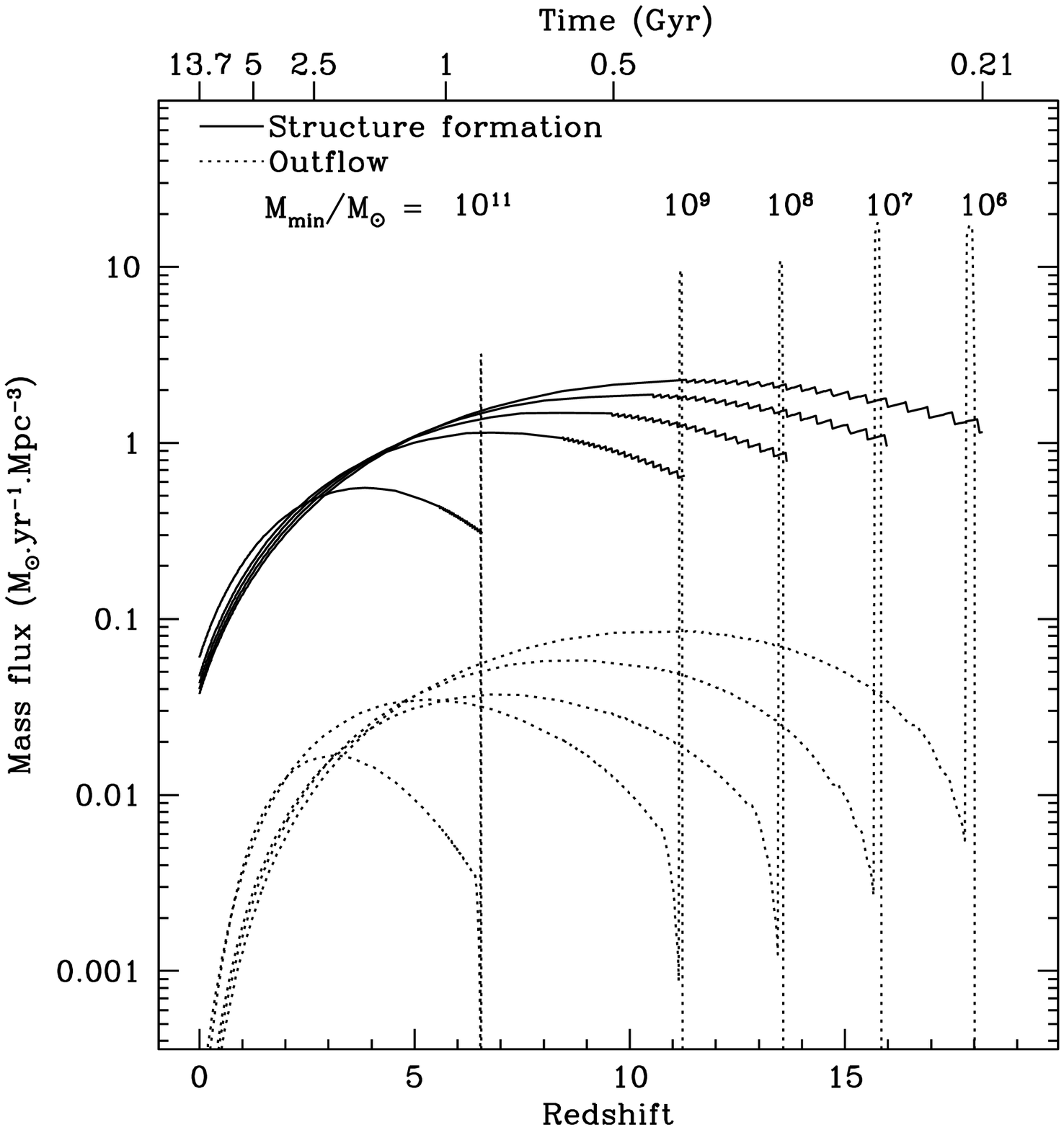}}
\end{center}
\caption{The rate of structure formation and outflow in Model 1. The mass flux corresponding to the structure formation is given by the solid curves and the rate of outflow from the structures to the IGM is shown by the dotted curves  as a function of redshift for all models shown in Figure~\ref{fig:model1SFRstarsZ}. } 
\label{fig:model1Outflow}
\end{figure}

The rate of accretion as well as the rate of outflow is shown in Figure~\ref{fig:model1Outflow}.
We see here that outflows from Pop III stars are very efficient at high redshift but their duration is 
very short due to the rapid increase in metallicity in the ISM. 
It is important to remember that the mass flux of outflows is very sensitive to $\epsilon$.
 As a consequence of the luminosity density obtained by the Sloan Digital Sky Survey (SDSS, \citet{blanton:03})  and by the 2dF galaxy redshift Survey \citep{croton:05b} together with the evaluation of the mass rate triggered by galactic winds  \citep{veilleux:05} we estimate that the actual mass flux of outflows  is of the order of 0.01 to 0.1  M$_\odot$/yr/Mpc$^3$. At $z = 2$, the results for Model 1
 show outflows at the lower end of this range.

\begin{figure}
\begin{center}
\resizebox{\textwidth}{!}{\includegraphics{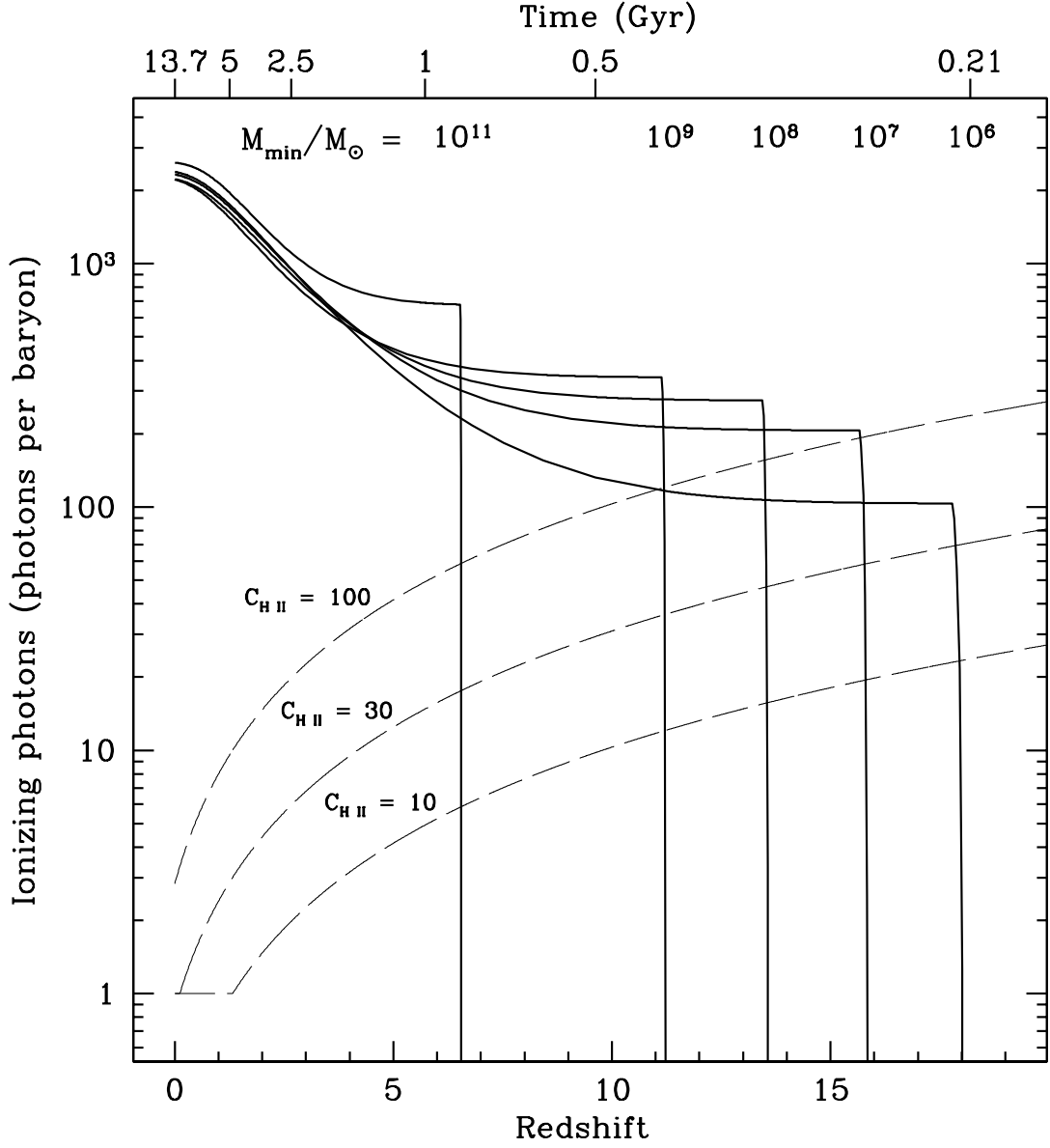}}
\end{center}
\caption{The cumulative number of ionizing photons emitted by stars is plotted as a function of redshift for all models of Figure~\ref{fig:model1SFRstarsZ}. The minimum number of photons per baryon necessary to fully reionize the IGM is plotted as a function of redshift by thin dashed lines for three different values of the clumpiness factor $C_\mathrm{H\ II}=10$, 30 and 100.}
\label{fig:model1Reionization}
\end{figure}

Among the chief motivating factors in developing a model of cosmic chemical evolution is
the early reionization of the Universe.  Several studies suggest
 that an early burst of star formation
as in Model 1, is sufficient \citep{venkatesan:03a,wyithe:03,venkatesan:03b,tumlinson:04,daigne:04}.
In Figure~\ref{fig:model1Reionization}, we show  the number of ionizing photons per baryon produced for each of our choices of $M_\mathrm{min}$. 
The procedure for  computing this stellar ionizing flux is explained in DOVSA. It is important to remember that only a fraction $f_\mathrm{esc}$ of these UV photons will escape the structures and therefore be available to ionize the IGM. The effective value of $f_\mathrm{esc}$ is poorly known but could vary from
about 1 to 30 \%.
The minimum number of photons required for complete reionization (see DOVSA) is also plotted for three possible clumpiness factors. The ionizing potential  clearly increases with $M_\mathrm{min}$ and decreasing redshift. The ratio of this minimum number (dashed line) to that for the stellar ionizing photons (solid line) gives the minimum fraction $f_\mathrm{esc}$ necessary to fully reionize the IGM.
It appears that all models are able to fully reionize the IGM for a clumpiness factor $C_\mathrm{H\ II} \sim 10$ and $f_\mathrm{esc} \sim 25 \% \to 1 \%$ for $M_\mathrm{min}=10^{6}\to 10^{11}\ \mathrm{M_\odot}$. When the clumpiness factor increases, only models with high $M_\mathrm{min}$ are still able to fully reionize the IGM with $f_\mathrm{esc} < 30 \%$. Typically, if $C_\mathrm{H\ II}\sim 30$ (100), we need $M_\mathrm{min} \ga 10^{7}\ \mathrm{M_{\odot}}$ ($10^{9}\ \mathrm{M_{\odot}}$). There is therefore a tendency to favor a high $M_\mathrm{min}$ and low reionization redshift if the clumpiness factor is high.

\begin{figure}
\begin{center}
\resizebox{\textwidth}{!}{\includegraphics{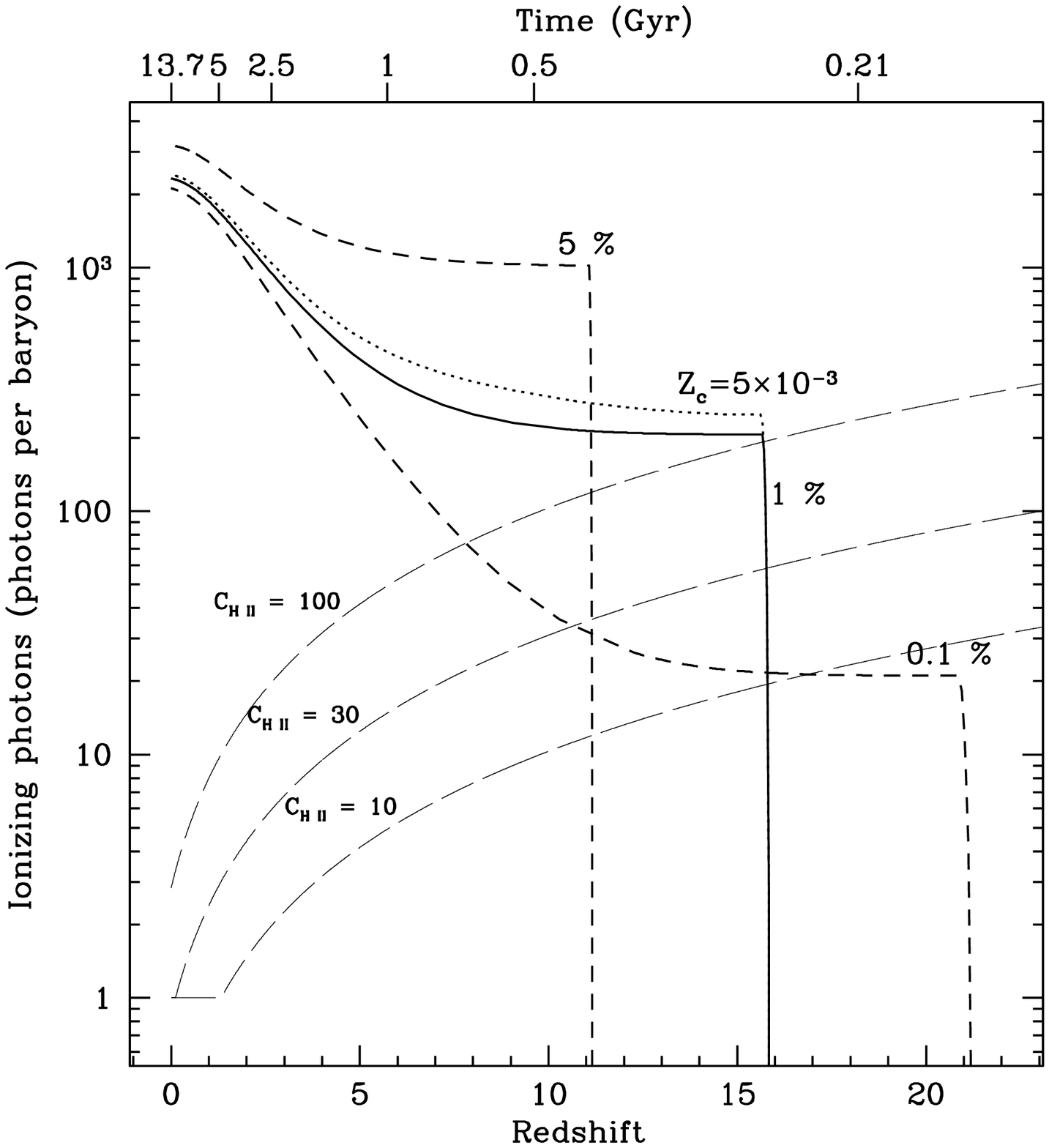}}
\end{center}
\caption{As in Figure~\ref{fig:model1Reionization} for $M_\mathrm{min}=10^{7}\ \mathrm{M_{\odot}}$, we plot the number of ionizing photons per baryon. The dotted curve corresponds to an increase of 
$Z_\mathrm{c}$ to $5 \times 10^{-3} Z_\odot$, ie. by a factor of 50. The dashed curves show the number of photons per baryon for two different choices for the initial baryon fraction in structures $0.1\ \%$ or $5\ \%$ as compared with the nominal choice of $1\ \%$.}
\label{fig:ZcRinit}
\end{figure}

In Figure~\ref{fig:ZcRinit}, we show the dependence of these results on the choice of 
$Z_\mathrm{c}$, the critical metallicity which shuts down massive star formation, and the initial baryon fraction in structures which allows star formation.  In all of the models discussed up to now, we have taken
$Z_\mathrm{c} = 10^{-4} Z_\odot$ and $f_\mathrm{b,struct}(z_\mathrm{init}) = 1\%$. We see in Figure~\ref{fig:ZcRinit} that results are almost
independent of $Z_\mathrm{c}$. Increasing $Z_\mathrm{c}$ leads to a modest change in the number of ionizing photons
per baryon.   The reason is that the first generation of stars formed
are in fact capable of producing a metallicity much larger than either  $10^{-4}$ or $5\times 10^{-3} Z_\odot$,
and massive star formation is rapidly terminated once these stars pollute the ISM. 
On the other hand, the results for the number of ionizing photons per baryon are quite sensitive to the choice of $f_\mathrm{b,struct}(z_\mathrm{init})$ as is the initial redshift for star formation. Our standard choice for $f_\mathrm{b,struct}(z_\mathrm{init})$ of 1\% corresponds to our conservative estimate of the minimum baryon fraction where sufficient dissipation occurs to allow star formation. Most galaxy formation simulations adopt a value of 10\% (cf. \citep{abadi:03}).

Decreasing $f_\mathrm{b,struct}(z_\mathrm{init})$ by a factor of 10 to 0.1\%
allows star formation to begin earlier (in this case at a redshift $z \sim 21$ for $M_\mathrm{min}=10^{7}\ \mathrm{M_\odot}$),
when the IGM is still quite dense, and reionization does not occur, even if $C_\mathrm{H\ II} = 10$.
If $f_\mathrm{b,struct}(z_\mathrm{init})$ is increased to 5\%, star formation occurs later ($z \sim 11$ for $M_\mathrm{min}=10^{7}\ \mathrm{M_{\odot}}$), when the IGM is less dense, and 
reionization occurs quite easily, even if $C_\mathrm{H\ II} = 100$ (only a fraction $f_\mathrm{esc} \sim 13 \%$ is needed in this case).

\begin{figure}
\begin{center}
\resizebox{\textwidth}{!}{\includegraphics{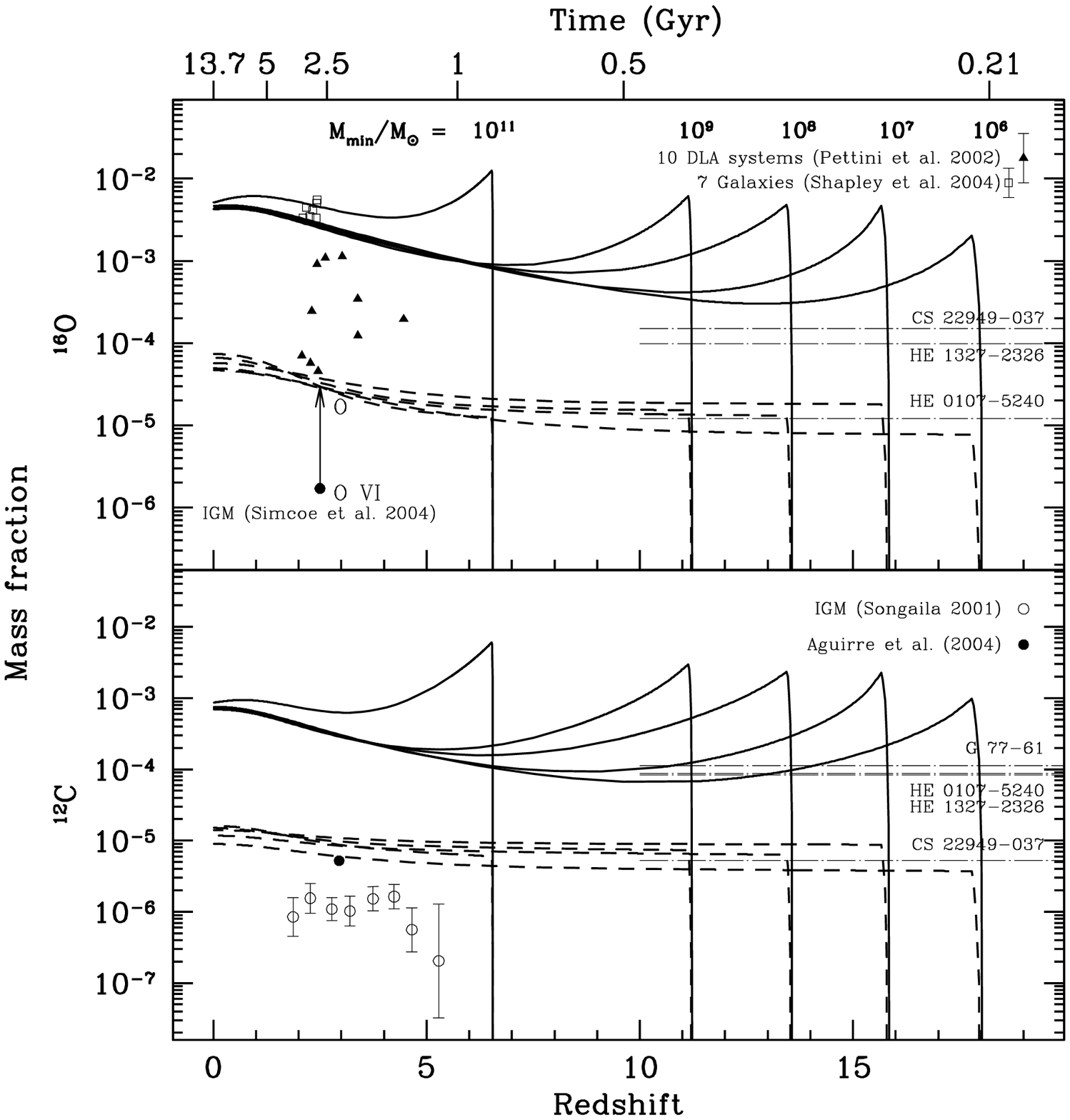}}
\end{center}
\caption{The evolution of the Oxygen and Carbon abundances  in Model 1. 
The mass fraction of Oxygen (top) and Carbon (bottom) as a function of redshift is plotted both in the ISM of structures (solid curves) and in the IGM (dashed curves) for all models of Figure~\ref{fig:model1SFRstarsZ}.
The observed abundances are taken from \citet{pettini:02} (filled triangles: Oxygen abundance in 10 DLAs), \citet{shapley:04} (open boxes: Oxygen abundance in 7 massive galaxies), \citet{simcoe:04} (Big dot: Oxygen VI in IGM. The arrow indicates the estimated value of the total Oxygen abundance),  \citet{songaila:01} (circle: Carbon IV abundance in the IGM) and \citet{aguirre:04} (big dot: total Carbon abundance in the IGM). Horizontal thin dashed lines indicate the measured abundances in the following four very metal-poor halo stars: CS 22949-037 (Oxygen and Carbon, from \citet{depagne:02}), HE 0107-2240 (Oxygen and Carbon, from \citet{bessell:04}), HE 1327-2326 (Oxygen and Carbon, from \citet{frebel:05}) and G 77-61 (Carbon, from \citet{plez:05}).}
\label{fig:model1OC}
\end{figure}

\begin{figure}
\begin{center}
\resizebox{\textwidth}{!}{\includegraphics{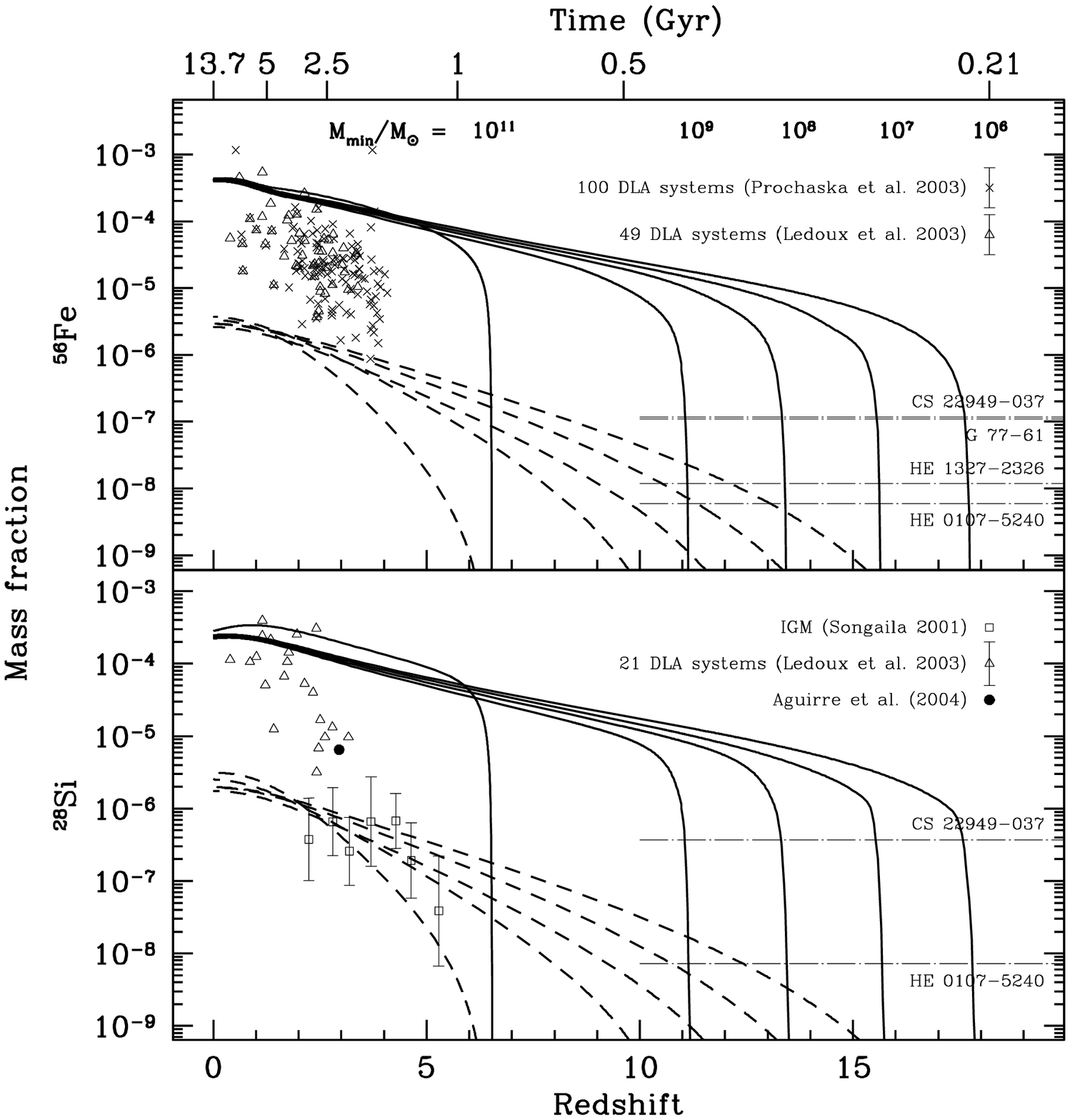}}
\end{center}
\caption{As in Figure~\ref{fig:model1OC}, the evolution of the Iron and Silicon abundances  in Model 1. Observed abundances are taken from \citet{prochaska:03} (crosses: Iron and Silicon abundances in 100 DLAs), \citet{ledoux:03} (triangles: Iron and Silicon abundances in 33 DLAs and sub-DLAs), \citet{songaila:01} (circle: Silicon IV abundance in the IGM) and \citet{aguirre:04} (big dot: total Silicon abundance in the IGM). Horizontal thin dashed lines indicate the measured abundances in the following four very metal-poor halo stars: CS 22949-037 (Iron and Silicon, from \citet{depagne:02}), HE 0107-2240 (Iron and Silicon, from \citet{bessell:04}), HE 1327-2326 (Iron, from \citet{frebel:05}) and G 77-61 (Iron, from \citet{plez:05}).
}
\label{fig:model1FeSi}
\end{figure}

Next, we show results for the evolution of C, O, Si, and Fe as a function of redshift in Model 1.
These are found in Figures~\ref{fig:model1OC} and \ref{fig:model1FeSi} (the yields vary with metallicity and come from  \citet{woosley:95}). Overall, one sees that the chemical abundances are well reproduced. In the cases of O and C, we see quite clearly the effects of the
hierarchical growth of structure.  After the initial burst of star formation by massive stars,
the ISM abundances within structures is diluted as metal-free baryons are accreted. 
As the normal mode of star formation begins to eject metals, these abundances subsequently begin to rise again.  Unfortunately, observations at sufficiently high redshift do not exist at present to test this
predicted feature. Because of the rapid ejection of metals in Model 1, we are able to achieve
the high abundances of C and O seen in the extremely iron-poor stars indicated in Figure~\ref{fig:model1OC}, in contrast to what is possible in Model 0, especially for C (see DOVSA).

Model 1 predictions in the IGM agree with the estimate of the oxygen abundance in the Lyman 
alpha forest (\citet{telfer:02}, \citet{simcoe:04}, \citet{aracil:04}, \citet{bergeron:05}). 
Note that only OVI is directly observed. Therefore, the estimate of the total O abundance is limited by the uncertainties of the ionization state of this element. We can reasonably consider that [O/H] can be as high as -2.5 (P. Petitjean, private communication). OVI was recently used to ascertain the distribution of metals in the IGM \citep{tumlinsonb:05}. Similarly, the IGM abundances of C agree
well with the determination by \citet{aguirre:04} and are (as expected) in excess  of the CIV
determinations of \citet{songaila:01,songaila:05}.  In contrast the  models currently fit
the SiIV abundances and underproduce the total Si
abundance based on the adopted yields of type II supernovae in Model 1.  
Indeed, we see from Fig. \ref{fig:model1FeSi} that the Si abundance derived 
from SiIV observations \citep{aguirre:04} is clearly underproduced by stars in the mass range
defined by Model 1. As we will see below in section 5.5, this translates in to a predicted value for
[Si/C] which is low compared with the determination of \citet{aguirre:04}.
Furthermore,
we note that as $M_\mathrm{min}$ is increased, it becomes more difficult to reproduce the IGM abundances.

\begin{figure}
\begin{center}
\resizebox{\textwidth}{!}{\includegraphics{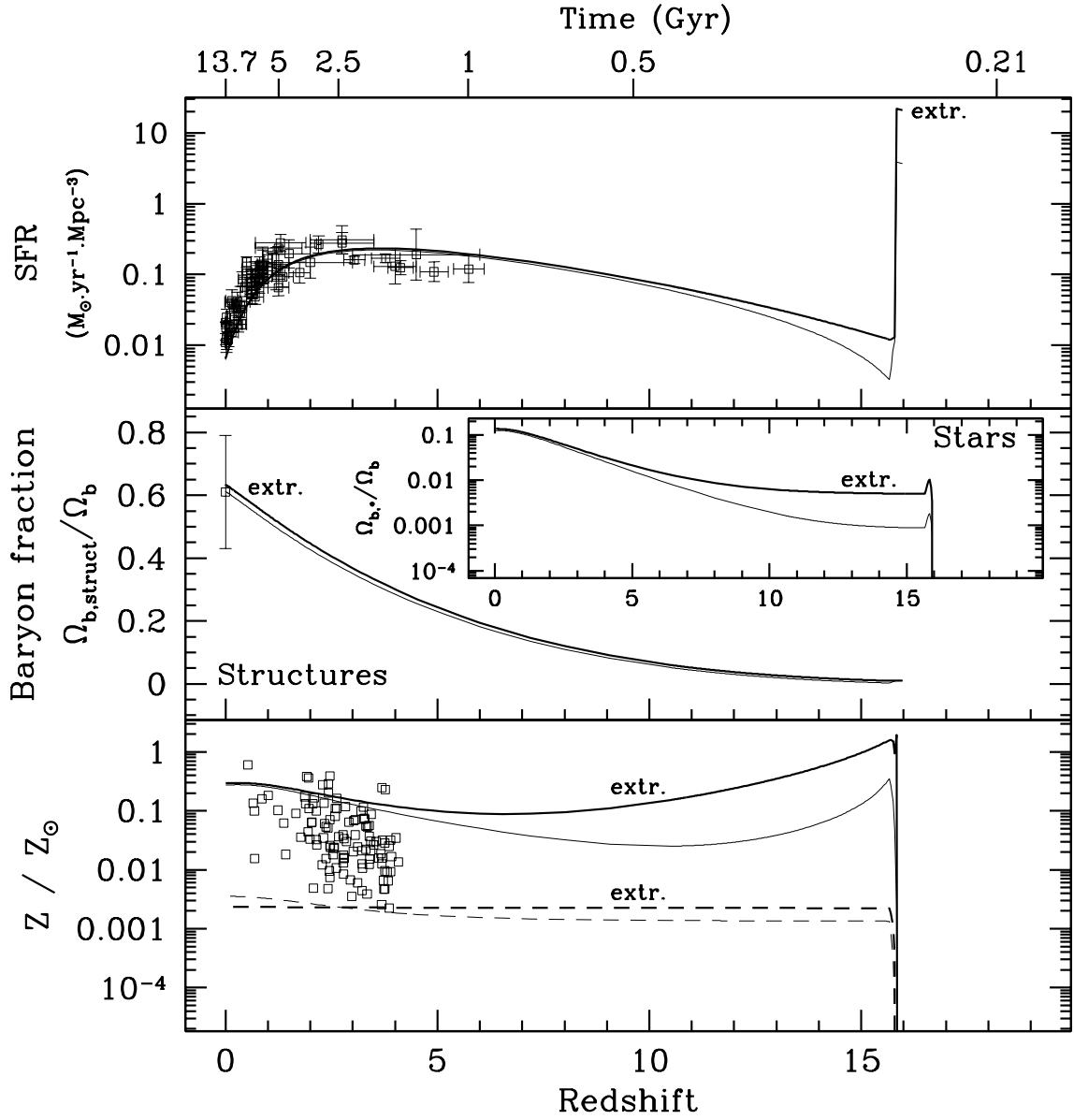}}
\end{center}
\caption{The star formation rate, baryon evolution and global metallicity for Models 1 (thin lines) and 1e (thick lines and labelled extr.) with $M_\mathrm{min}=10^{7}\ \mathrm{M_{\odot}}$.}
\label{fig:extremModel1SFRFracBZ}
\end{figure}

To test the importance of the normal mode of star formation on the evolution of the chemical
abundances, we also consider a set of models in which the efficiency of outflow
is reduced, thereby increasing the relative contribution 
of Population III stars to the overall metallicity in the IGM.
To compensate for the reduction in outflow, the SFR of the massive mode is increased
(cf. Model 1e in Table 2) which enhances the ability of the population to reionize the IGM.  
In Figure~\ref{fig:extremModel1SFRFracBZ}, 
we show the SFR, baryon fraction, and metallicity for Model 1e compared with the analogous result for Model 1, restricting attention to  $M_\mathrm{min} = 10^7$ M$_\odot$.  With the exception of 
$\epsilon$, other Model 0 parameters are left unchanged.  The difference in the SFR between Models 1 and 1e is predominantly in the initial burst of massive star formation.  The massive SFR is typically 5 times larger initially, while at lower redshifts the SFR,
which is determined by $\nu_1$ is identical.  Similarly, the baryon fraction in stars is the same for both 
Models 1 and 1e, but 1e exhibits a strong peak (close to $f_b^* = 1$), which 
later relaxes due to hierarchical growth.  Unfortunately it is not possible to observationally
test for these differences.  In contrast, however, the overall metallicity does present us
with a real test (requiring suitably accurate data) between Models 1 and 1e.
In Model 1, the IGM metallicity is clearly changing with redshift.  For example, 
when $M_\mathrm{min} = 10^7$ M$_\odot$, the metallicity increases by a factor of about 20 from
a redshift of 16 to 0 and by a factor of about 2 from a redshift of 5 to 0.  In Model 1e, 
the evolution of $Z$ is very nearly constant in the IGM, with its final value fixed at very high redshift.

\begin{figure}
\begin{center}
\resizebox{\textwidth}{!}{\includegraphics{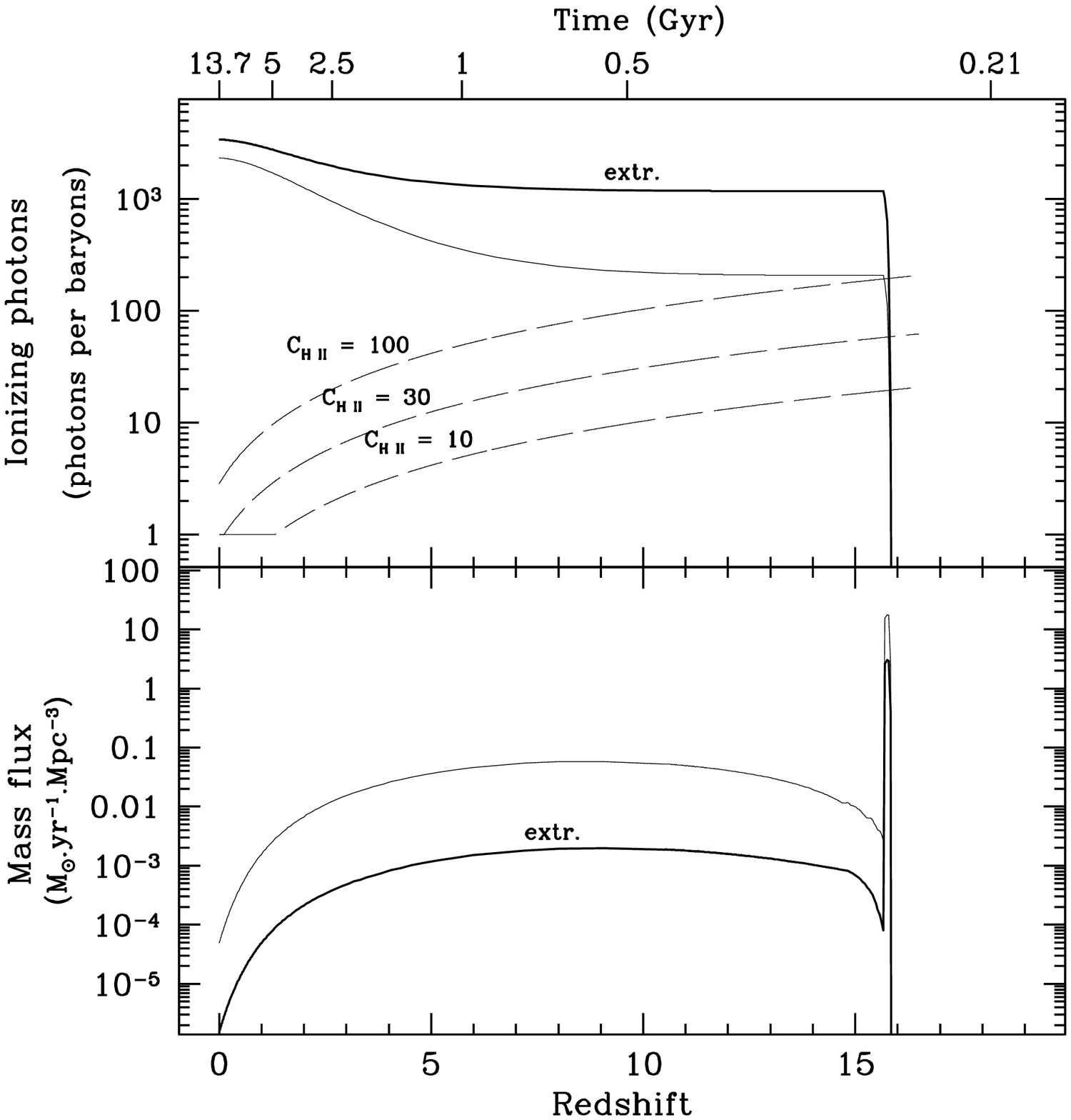}}
\end{center}
\caption{The reionization potential and outflow rate for Models  1 (thin lines) and 1e (thick lines and labelled extr.) with $M_\mathrm{min}=10^{7}\ \mathrm{M_{\odot}}$.}
\label{fig:extremModel1FluxOutflow}
\end{figure}

One of the key benefits of Model 1e is its ionizing potential relative to Model 1.
In Figure~\ref{fig:extremModel1FluxOutflow}, we compare the number of ionizing photons per baryon in Models 1 and 1e.  
As one can see, whereas Model 1 (with $M_\mathrm{min} = 10^7$ M$_\odot$) provided the minimum number of ionizing photons when $C_\mathrm{H\ II}=30$, Model 1e produces a factor of 5 times more
ionizing photons per baryon at high redshift.  

As explained earlier, to achieve a high ionization efficiency, we increased the SFR parameter
$\nu_2$. However, in order to avoid  overproduction of IGM metals formed by Pop III stars, 
 we are required to adjust $\epsilon$ downwards. Consequently, while outflow is very efficient at very high redshift, it is suppressed at later times due to hierarchical growth and is unable to eject metals coming from subsequent structures. The mass flux of outflows in Models 1 and 1e are compared in
Figure~\ref{fig:extremModel1FluxOutflow}.  Recall that observations indicate a flux of
0.01 -- 0.1 M$_\odot$/yr/Mpc$^3$. 
We see that the predicted outflow in Model 1 is ten times higher than that in Model 1e. 
Model 1 seems to be in better agreement with the data than Model 1e, which is probably insufficient to fit the high outflows observed in galaxies at $z = 2$ (see also \citet{bertone:05,rupke:05}).

\begin{figure}
\begin{center}
\resizebox{\textwidth}{!}{\includegraphics{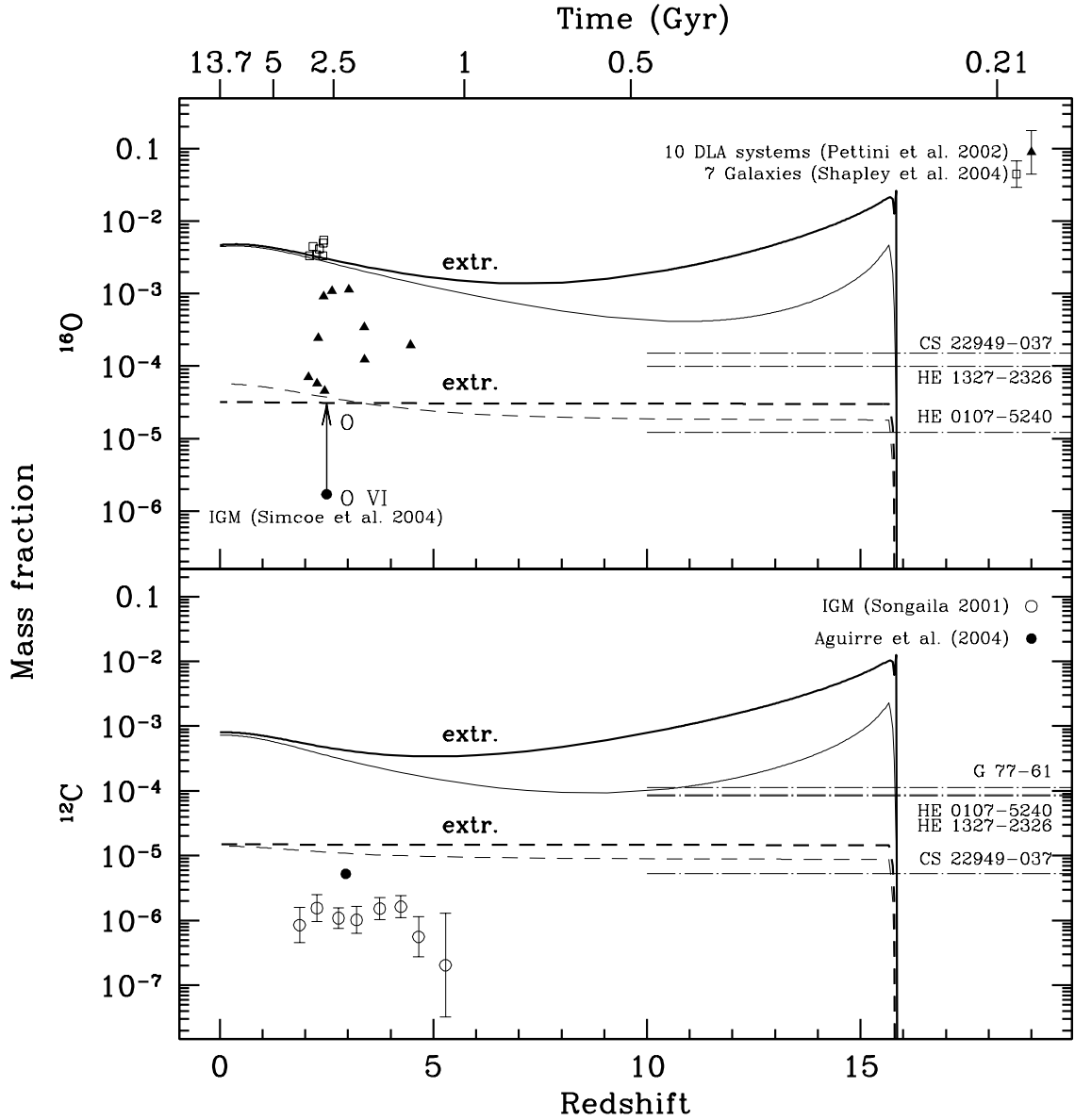}}
\end{center}
\caption{The evolution of the O and C abundances in Models 1 (thin lines) and 1e (thick lines and labelled extr.) with $M_\mathrm{min}=10^{7}\ \mathrm{M_{\odot}}$.}
\label{fig:extremModel1OC}
\end{figure}

\begin{figure}
\begin{center}
\resizebox{\textwidth}{!}{\includegraphics{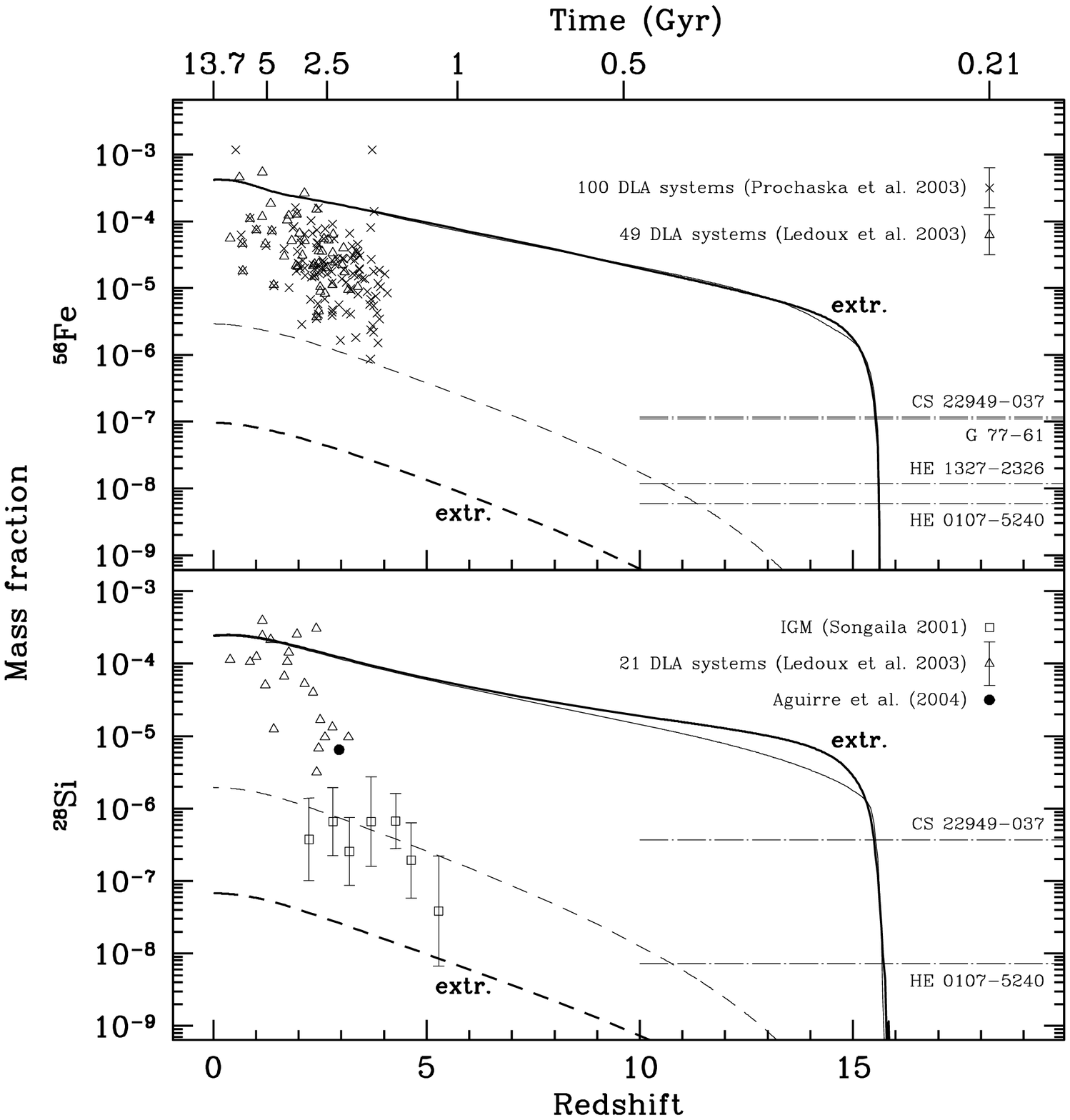}}
\end{center}
\caption{The evolution of the Si and Fe abundances in Model 1 (thin lines) and 1e (thick lines and labelled extr.) with $M_\mathrm{min}=10^{7}\ \mathrm{M_{\odot}}$.}
\label{fig:extremModel1FeSi}
\end{figure}

Despite the advantage in ionization potential, it is also difficult to reconcile  
the chemical history of Model 1e  with IGM observations.
In Figures~\ref{fig:extremModel1OC} and \ref{fig:extremModel1FeSi}, 
we compare the evolution of C, O, Si, and Fe in Models 1 and 1e.  The early burst of  
star formation in Model 1e produces a prompt initial enrichment  of the IGM
in C and O, which show very little evolution at lower redshifts.  In contrast, the evolution of Si and Fe
is particularly poor, as this models fails to reproduce the observed IGM abundances of these
elements at low redshift.  

\subsection{Model 2a}

\begin{figure}
\begin{center}
\resizebox{\textwidth}{!}{\includegraphics{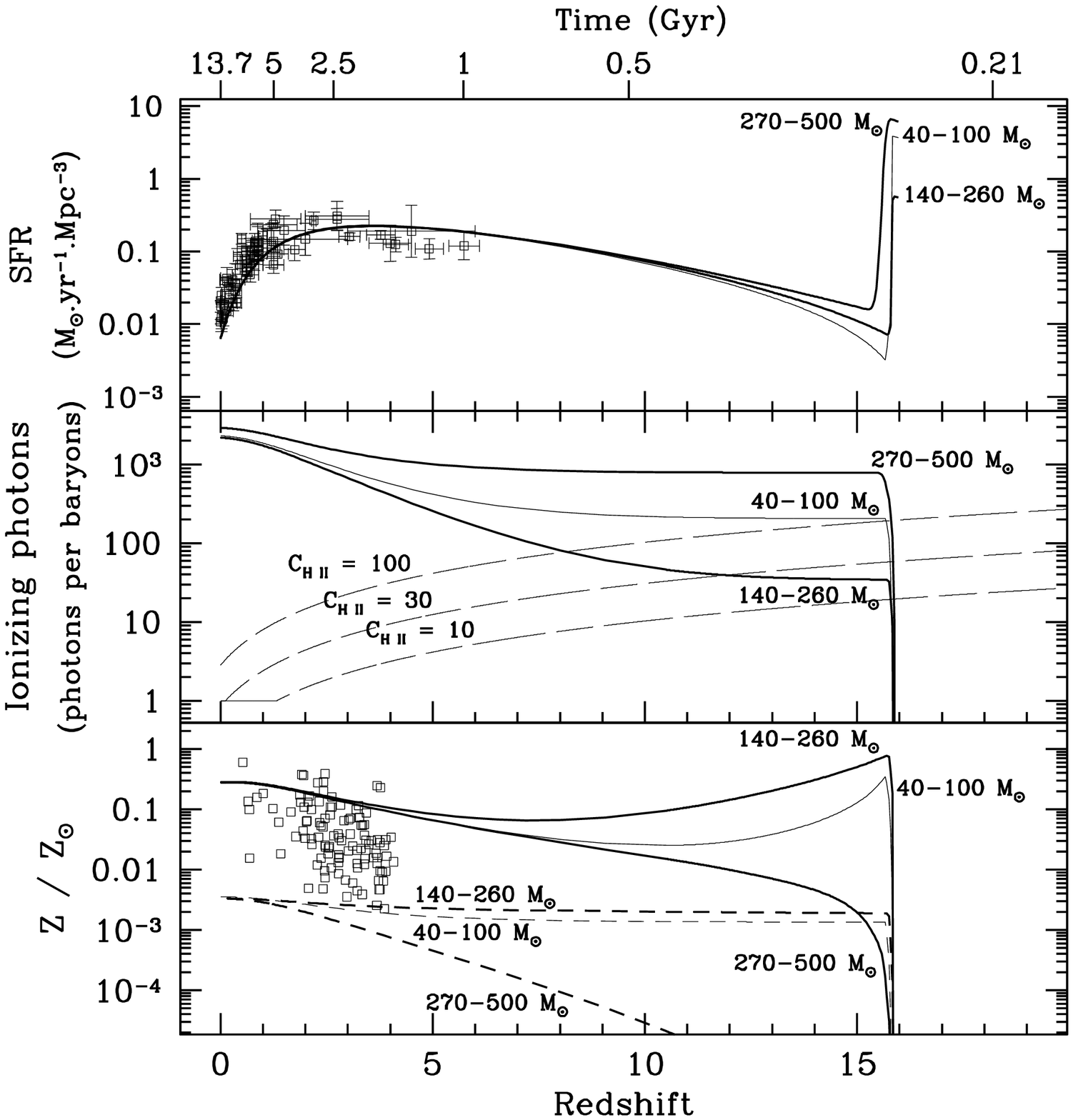}}
\end{center}
\caption{The star formation rate, reionization potential, and global metallicity for Models 1, 2a, and 2b with $M_\mathrm{min}=10^{7}\ \mathrm{M_{\odot}}$. }
\label{fig:model2SFRZ}
\end{figure}

Next we consider a massive mode made of stars with masses in the range 140 -- 260 M$_\odot$
which explode as pair-instability supernovae. In Figure~\ref{fig:model2SFRZ}, we compare the star formation rate, ionization potential and metallicity  in Model 2a with that of Model 1. 
In Model 2a, the initial rapid burst of star formation leads to large metallicities 
at high redshift.  In the ISM, these are diluted by hierarchical growth. In the IGM, where there is no dilution, the metallicity remains relatively flat.  
Our results confirm that PISN models are less efficient at reionization than
a more standard IMF as in Model 1. Indeed, even for a clumpiness factor, 
$C_\mathrm{H\ II} \sim 10$, Model 2a would require a photon escape fraction in excess of 
50\% placing this model in a difficult position to explain the early reionization of the IGM.

\begin{figure}
\begin{center}
\resizebox{\textwidth}{!}{\includegraphics{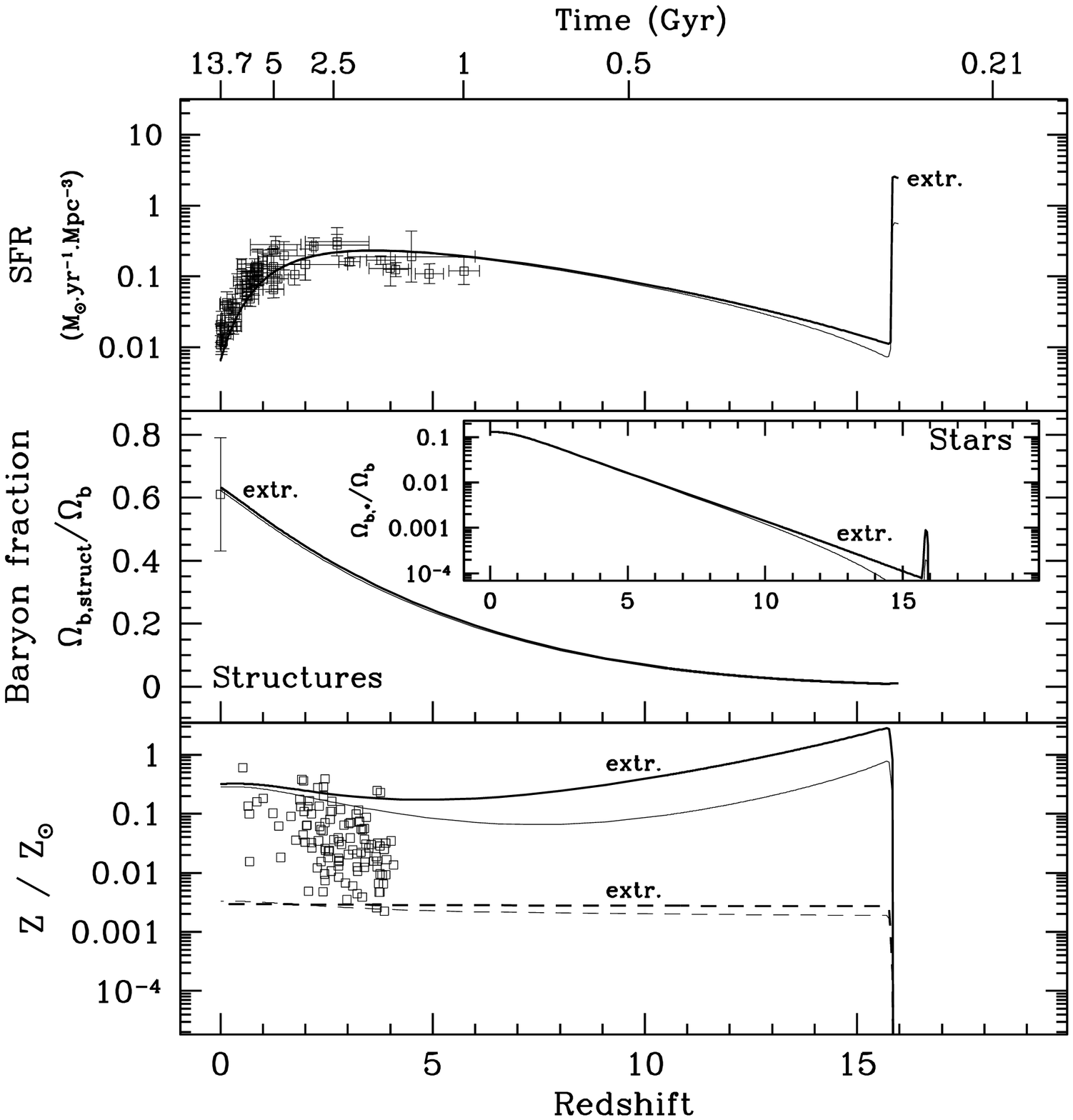}}
\end{center}
\caption{As in Figure~\ref{fig:extremModel1SFRFracBZ} for Model 2a.}
\label{fig:extremModel2aSFRFracBZ}
\end{figure}

We have also considered a more extreme version of Model 2a (2ae) in which approximately 90\%
of the IGM oxygen abundance at $z = 2.5$ originates in Pop III. As  with the relation between Models 1 and 1e, Model 2ae has a star formation rate characterized by $\nu_2$ which is about 5 times larger than Model 2a. This enhancement is seen in the early SFR is shown in Figure~\ref{fig:extremModel2aSFRFracBZ} where the evolution of the baryon density and overall metallicity
are also compared with Model 2a.

\begin{figure}
\begin{center}
\resizebox{\textwidth}{!}{\includegraphics{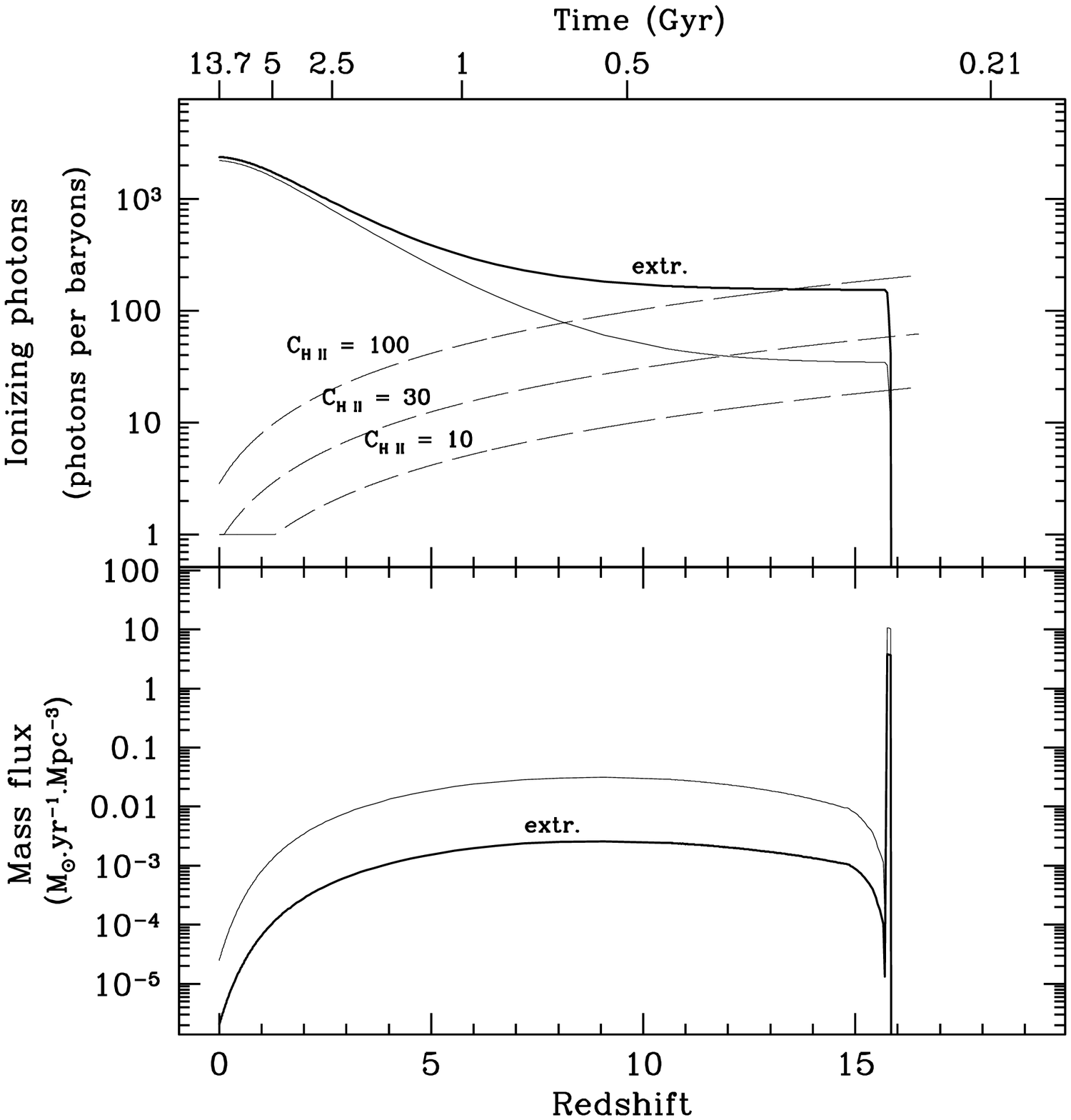}}
\end{center}
\caption{As in Figure~\ref{fig:extremModel1FluxOutflow} for Model 2a.}
\label{fig:extremModel2aFluxOutflow}
\end{figure}

In the more extreme case, (Model 2ae), 
the number of ionizing photons per baryon (shown in Figure~\ref{fig:extremModel2aFluxOutflow})
is increased to about 100 at $z \sim 16$, making
this version of the PISN model acceptable based on reionization.  
However, as was the case in Model 1e, the mass flux of outflows at low redshift is
very small in Model 2ae.

\begin{figure}
\begin{center}
\resizebox{\textwidth}{!}{\includegraphics{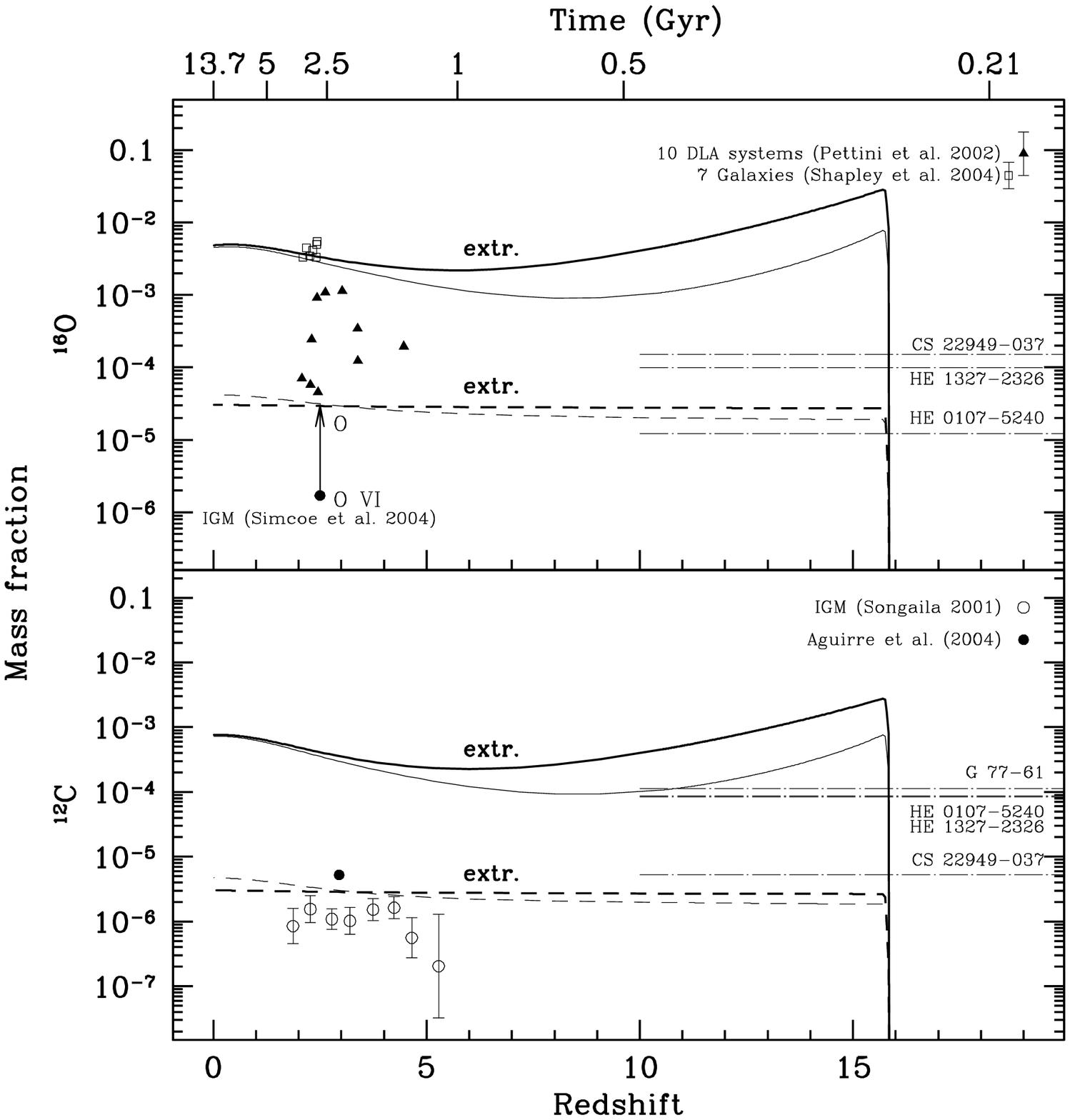}}
\end{center}
\caption{As in Figure~\ref{fig:extremModel1OC} for Model 2a.}
\label{fig:extremModel2aOC}
\end{figure}

\begin{figure}
\begin{center}
\resizebox{\textwidth}{!}{\includegraphics{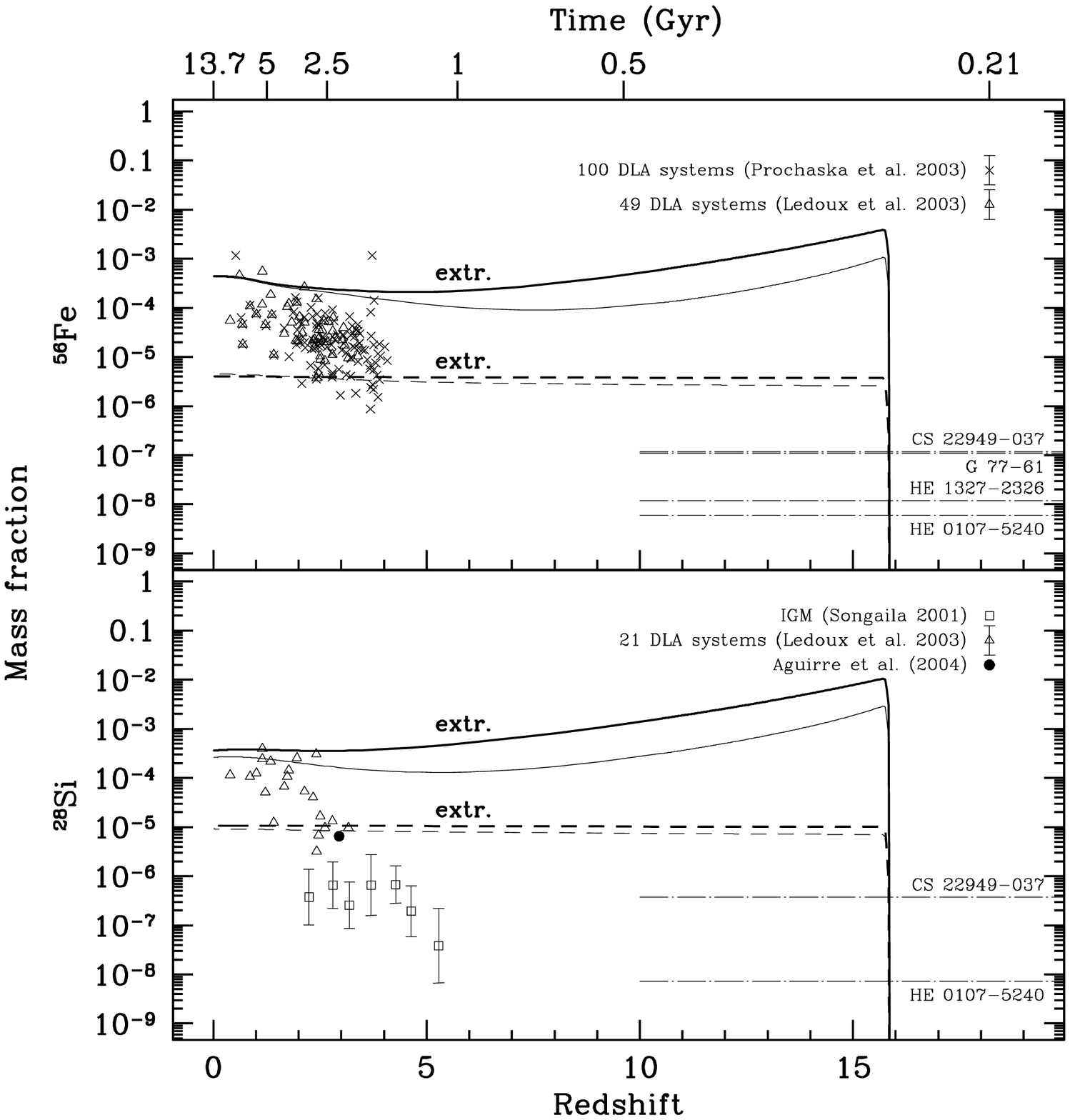}}
\end{center}
\caption{As in Figure~\ref{fig:extremModel1FeSi} for Model 2a.}
\label{fig:extremModel2aFeSi}
\end{figure}

The evolution of C, O, Si, and Fe are shown in Figures~\ref{fig:extremModel2aOC} and \ref{fig:extremModel2aFeSi}.
This model is acceptable for all nuclei (see however section 5.5 for a comparison of abundance
ratios with the those observed in iron-poor stars),
and  predicts little evolution in the IGM. 

\subsection{Model 2b}

Finally, we turn to the case of Model 2b, where the massive mode is made of stars with masses in the range 270 -- 500 M$_\odot$ which collapse directly to black holes without any ejection of heavy elements into either the ISM or the IGM.   The star formation rate, ionization potential and metallicity 
in Model 2b are also shown in Figure~\ref{fig:model2SFRZ}. As one can see, the SFR and the number of
ionizing photons per baryon are quite high relative to Models 1 and 2a.  Because these stars do 
not produce any metals, the chemical evolution of this model is identical to that of Model 0.
This is mostly problematic in regard to the very iron-poor stars whose abundances could not
be explained in this case.

\subsection{Abundance Ratios}

Before concluding, we examine and compare a number of abundance ratios 
in Models 1 and 2a.  In particular, we consider the evolution of O/Fe, C/Fe, and Si/C.
These ratios are of particular interest when trying to model the very iron-poor stars
which exhibit anomalously high ratios of O/Fe, C/Fe and to some extent Si/C. The latter is of
potential interest at lower redshifts where some derivations of the IGM abundance of Si/C
are relatively high coming from SiIV and CIV observations.

\begin{figure}
\begin{center}
\resizebox{\textwidth}{!}{\includegraphics{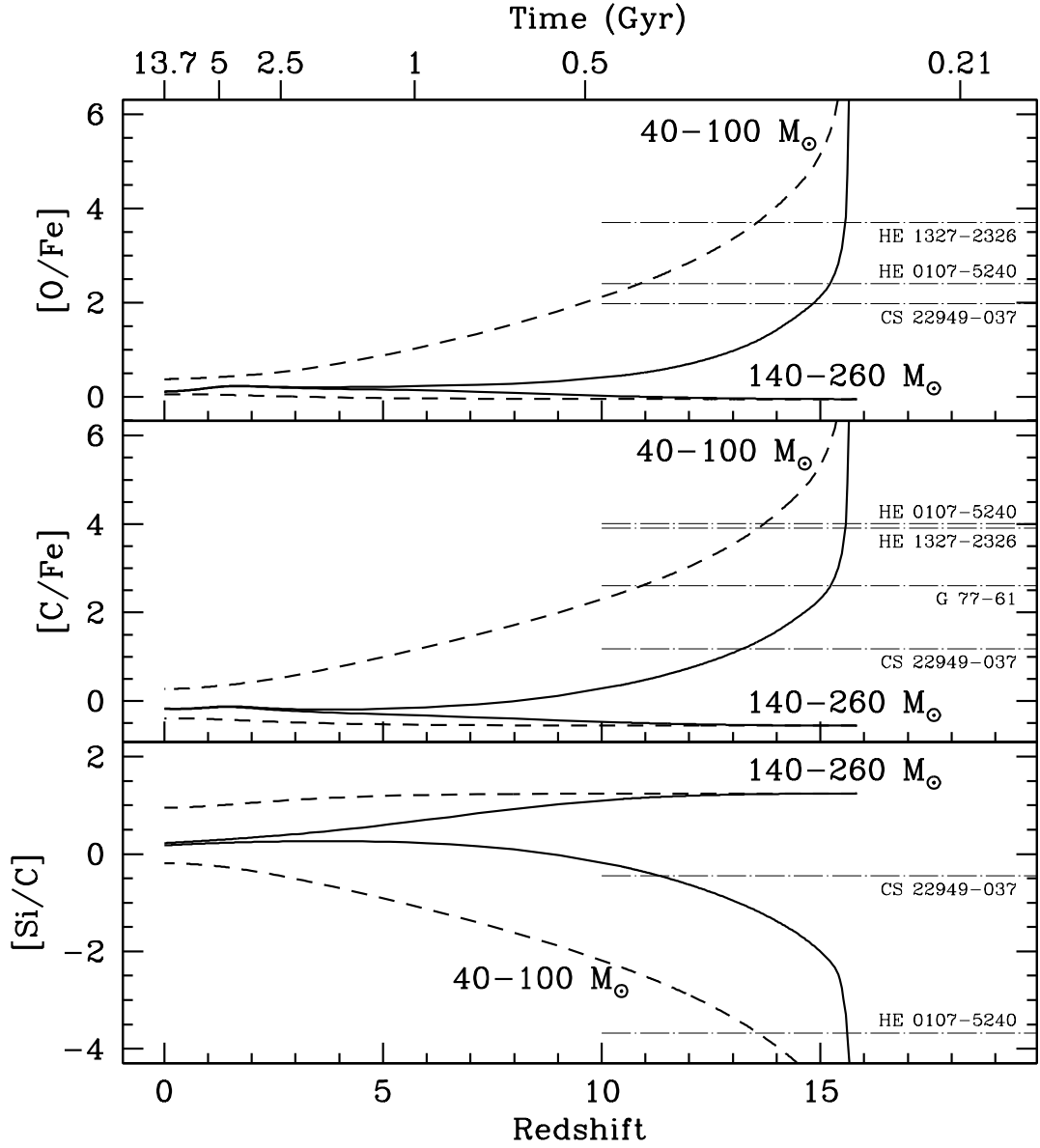}}
\end{center}
\caption{The evolution of the [O/Fe], [C/Fe] and [Si/C] ratios in Models 1 and 2a. The ratios are plotted for two mass ranges of the massive mode: 40-100 $\mathrm{M_{\odot}}$ (Model 1) and 140-260 $\mathrm{M_{\odot}}$ (Model 2a). As in the previous figures, the solid line (dashed line) stands for the ISM (IGM). Horizontal thin dashed lines indicate the measured abundances in the following four very iron-poor halo stars: CS 22949-037 \citep{depagne:02}, HE 0107-2240 \citep{bessell:04}, HE 1327-2326 \citep{frebel:05} and G 77-61 \citep{plez:05}.}
\label{fig:model1OFeCFeSiC}
\end{figure}

As shown in Figure~\ref{fig:model1OFeCFeSiC}, the extremely iron-poor stars observed over the
last several years show values of [O/Fe] between 2-4, which is significantly higher than
typical Population II values which range between 0.5 and 1.  A similar pattern is seen for [C/Fe]
which is also very large in these stars.  Model 1 reproduces these patterns quite well because
high C,O/Fe ratios are expected in massive 
type II supernova explosions.  The Si/C is also suppressed, and the fit to HE 0107-5240
is also quite good whereas the model somewhat underpredicts the Si/C ratio for CS 22949-037.
This result reinforces the notion that these stars are indeed very old and reveal the
composition of the gas at the end of Population III.
In contrast, Model 2a and PISN are not capable of achieving the element ratios observed in these stars.

At low redshift, one can attempt to use the IGM abundances of Si/C to try to differentiate between models. From Figure~\ref{fig:model1OFeCFeSiC}, we see that Model 1 predicts a [Si/C] ratio of order
-0.25 whereas in Model 2 it is closer to 1 at a redshift between 2 and 4. 
This is due to the large
Si yields found in PISN models \citep{heger:02}. 
There have been some recent observations of ionized Si and C 
\citep{songaila:01, schaye:03,aguirre:04}
from which a model-dependent ratio was inferred \citep{hm:96}.  Indeed the result is highly dependent on the assumption of the UV background model.  The most extreme case is that of a UV background
powered solely quasars. In this case, [Si/C] = 1.48, far exceeding the model predictions found here.
When a quasar plus galaxy model is considered, the ratio drops to 0.77, which is close to the result
of Model 2a. When the UV flux is softened at high redshift, the ratio drops to 0.48.  
This ratio may present a serious challenge to Model 1 which clearly underproduces Si.
We note however, that the resulting Si abundance is sensitive to several parameter choices.
For example, lowering $m_{inf}$ to 20 M$_\odot$ results in more Si (but aversely affects
the C,O/Fe ratios as shown in DOVSA. Increasing the slope of the IMF and more importantly
decreasing the massive mode astration factor, $\nu_2$, both lead to enhanced Si production.
The latter effect can be seen in Fig. \ref{fig:extremModel1FeSi}, where we show the effect of 
increasing $\nu_2$ by a factor of 5 leads to decrease in Si production by a factor of about 30,
with little change in C.
While it is tempting to argue for a model such as 2a on
the basis of [Si/C] \citep{qian:05}, given the large uncertainty inherent in the UV model, 
it is not clear which model (early star formation and chemical evolution or the UV background model) 
is being tested by these observations. The total IGM element abundances inferred from ionized values remain uncertain. Nevertheless, Model 1 currently predicts a value of [Si/C] which is too small
and may signal the need for the presence of some component of more massive stars such as in Model 2a.


\section{Conclusions} 
\label{sec:conclusions}

We have  calculated the
cosmic star formation history corresponding to different minimum masses for 
the initial halo structures, 
spanning 
$10^{6}$ to $10^{11}\ \mathrm{M_{\odot}}$. We include 
realistic gas outflows from the structures, 
powered by the kinetic energy of supernovae and which take into account the increase of the escape velocity due to the growth of structure. 
 We then deduce  the baryon content and the chemical
composition of the structures and of the intergalactic medium (IGM).
We show that  
with a minimum halo mass of $10^{7}$--$10^{8}\
\mathrm{M}_{\odot}$  and 
a moderate outflow efficiency , we are able to
 reproduce both the fraction of baryons in the
structures at the present time and the early chemical enrichment of the
IGM. The intensity of the formation rate of ``normal'' stars is also
well constrained by the observations: it has to be dominated by 
star formation in elliptical galaxies, except perhaps  at very low
redshift. The fraction of baryons in stars is also predicted as are also 
the type Ia and II supernova event
rates. The comparison with SN observations in the redshift range 
$z=0-2$ allows us to set  strong constraints on the time delay of type Ia
supernovae (a total delay of $\sim$4 Gyr is required to fit the data), on the lower end of the
mass range of the progenitors (2 - 8 $\mathrm{M}_{\odot}$) and on the
fraction of white dwarfs that reproduce the  type Ia supernova (about 1 per
cent). 
The type II supernova rate is also well fitted in the same
range of redshifts in our models and it is directly correlated to the cosmic SFR. 
We incorporate an improved treatment of
structure formation compared to our previous work \citep{daigne:04}
 that  leads to  new insights into the initial mass function
(IMF) of the population III stars at high redshift. 
We compare three
possible mass ranges: 40-100 $\mathrm{M}_{\odot}$ (normal
supernovae), 140-260 $\mathrm{M}_\odot$ (pair-instability supernovae)
and 270-500 $\mathrm{M_{\odot}}$. 

We have demonstrated that the fraction in the initial starburst of zero  metallicity  stars below
270 $\mathrm{M_{\odot}}$
must be 
limited in order to avoid premature overenrichment of the IGM. Specifically, we predict
that about 10 - 20 \% of the metals present in the IGM  at $z = 0$ have been produced
by population III stars at very high $z$. The remaining 80 - 90 \% are ejected later  by
galaxies forming normal stars, with a maximum efficiency of the outflow
occurring at a redshift of about 5.
In full agreement with \citet{daigne:04},
because of the chemical constraints, including both for the IGM and very metal-poor halo stars,
$10^{-3}$ of the baryons must lie in the first massive stars  
in order to   produce  enough ionizing photons
to  allow early reionization of the IGM by  $z\sim 15$. 
The case of the very massive mass range (270-500 $\mathrm{M}_{\odot}$) is highly
efficient regarding the ionizing flux but cannot reproduce alone the
global chemical evolution. 
A massive component of stars with masses in the 40 -- 100  $\mathrm{M}_{\odot}$
account for the observed ISM and IGM abundances with the possible exception of IGM Si.
While more massive stars in the range 140 -- 260  $\mathrm{M}_{\odot}$ produce ample amounts of Si, 
they can not produce the C,O/Fe ratios observed in extremely iron-poor stars, nor are they able to 
efficiently reionize the IGM at high redshift.  If future observations bear out a high Si/C ratio in the 
IGM, we would be led to consider a hybrid model of massive stars.

In summary, we have demonstrated the sensitivity of chemical evolution to different  constraints, including the minihalo 
mass range, the role of outflows and the redshift of structure formation, correlated with different ranges of massive stars.
 We have evaluated the rates as a function of epoch of both SNII and SNIa. We conclude that only 10 to 20  \%  of the metals in the IGM are produced by Population III.
Addition of a massive star component to Population III with masses in the conventional stellar mass range provides an effective means  of simultaneously accounting for both early reionization and the chemical evolution both of the oldest  extreme metal-poor stars and of the IGM.

\acknowledgements
The authors gratefully thank  Yannick Mellier and Patrick Petitjean for frequent  discussions and Adrian Jenkins for giving us the code computing the dark matter mass functions. The work of K.A.O., F.D. and E.V. was supported by the Project "INSU - CNRS/USA", and the work of K.A.O. was also supported partly by DOE grant
DE--FG02--94ER--40823.

\bibliographystyle{apj}
\bibliography{dossv4}

\end{document}